\documentclass{amsart}
\usepackage{amssymb}
\usepackage{amsmath}
\usepackage{graphicx}
\usepackage{amscd}

\newtheorem{theorem}{Theorem}
\theoremstyle{plain}

\newtheorem{lemma}{Lemma}

\numberwithin{equation}{section}

\begin{document}
\title[double indexed Hill estimator]{A double-indexed functional Hill
process and applications}

\author{Gane Samb LO$^{*}$}
\author{Modou Ngom$^{**}$}
\address{$^{*}$ LSTA, Universit\'e Pierre et Marie Curie, France and LERSTAD, Universit\'e Gaston Berger
de Saint-Louis, SENEGAL\\
gane-samb.lo@ugb.edu.sn, ganesamblo@ganesamblo.net}
\address{$^{**}$ LERSTAD, Universit\'e Gaston Berger de Saint-Louis, SENEGAL%
\\
ngomodoungom@gmail.com}

\begin{abstract}
Let $X_{1,n} \leq .... \leq X_{n,n}$ be the order statistics associated with
a sample $X_{1}, ...., X_{n}$ whose pertaining distribution function (%
\textit{df}) is $F$. We are concerned with the functional asymptotic
behaviour of the sequence of stochastic processes 
\begin{equation}
T_{n}(f,s)=\sum_{j=1}^{j=k}f(j)\left( \log X_{n-j+1,n}-\log
X_{n-j,n}\right)^{s} ,  \label{fme}
\end{equation}
indexed by some classes $\mathcal{F}$ of functions $f:\mathbb{N}%
^{\ast}\longmapsto \mathbb{R}_{+}$ and $s \in ]0,+\infty[$ and where $k=k(n)$
satisfies 
\begin{equation*}
1\leq k\leq n,k/n\rightarrow 0\text{ as }n\rightarrow \infty .
\end{equation*}

\noindent We show that this is a stochastic process whose margins generate
estimators of the extreme value index when $F$ is in the extreme domain of
attraction. We focus in this paper on its finite-dimension asymptotic law
and provide a class of new estimators of the extreme value index whose
performances are compared to analogous ones. The results are next
particularized for one explicit class $\mathcal{F}$.
\end{abstract}

\subjclass{Extreme values theory; Asymptotic distribution; Functional
Gaussian and nongaussian laws; Uniform entropy numbers; Asymptotic
tightness, Stochastic process of estimators of extremal index; Slowly and
regularly varying functions}.

\maketitle

\section{Introduction} \label{sec1}

\subsection{General introduction}

In this paper, we are concerned with the statistical estimation of the
univariate extreme value index of a \textit{df} $F$, when it is available. But rather than doing this by one statistic, we are going to use
a stochastic process whose margins generate estimators of the extreme value
index (SPMEEXI). To precise this notion, let $X_{1},X_{2},...$ be a sequence
of independent copies (s.i.c) of a real random variable ($rv$) $X>1$ with 
\textit{df} $F(x)=\mathbb{P}(X\leq x)$. $F$ is said to be in the extreme
value domain of attraction of a nondegenerate \textit{df} $M$ whenever there
exist real and nonrandom sequences \ $(a_{n}>0)_{n\geq 1}$ and $%
(b_{n})_{n\geq 1}$ such that for any continuity point $x$\ of $M,$%
\begin{equation}
\lim_{n\rightarrow \infty }P(\frac{X_{n,n}-b_{n}}{a_{n}}\leq
x)=\lim_{n\rightarrow \infty }F^{n}(a_{n}x+b_{n})=M(x).  \label{dl05}
\end{equation}%
It is known that $M$ is necessarily of the family of the Generalized Extreme
Value (GEV) \textit{df} : 
\begin{equation*}
G_{\gamma }(x)=\exp (-(1+\gamma x)^{-1/\gamma })\text{, }1+\gamma x\geq 0,
\end{equation*}%
parameterized by $\gamma \in \mathbb{R}$. The parameter $\gamma $ is called
the extreme value index. There exists a great number of estimators of $%
\gamma $, going back to first of all of them, the Hill's one defined by 
\begin{equation*}
T_{n}(f,s)=k^{-1}\sum_{j=1}^{k}j(\log X_{n-j+1,n}-\log X_{n-j,n}),
\end{equation*}%
where for each $n$, $k=k(n)$ is an integer such that 
\begin{equation*}
1\leq k\leq n,\text{ }k\rightarrow \infty ,\text{ }k/n\rightarrow 0\text{ as 
}n\rightarrow \infty .
\end{equation*}%
A modern and large account of univariate Extreme Value Theory can be found
in Beirlant, Goegebeur and Teugels \cite{bgt}, Galambos \cite{galambos}, de
Haan \cite{dehaan} and \cite{dehaan1}, Embrechts \textit{et al.} \cite%
{embrechts} and Resnick \cite{resnick}. One may estimate $\gamma $ by one
statistic only. This is widely done in the literature. But one also may use
a stochastic process of statistics $\{T_{n}(f),f\in \mathcal{F}\}$ indexed
by $\mathcal{F}$, such that for any fixed $f\in \mathcal{F},$ there exists a
sequence of nonrandom and positive real coefficients $(a_{n}(f))_{n\geq 1}$
such that $T_{n}^{\ast }(f)=T_{n}(f)/a_{n}(f)$ is an asymptotic estimator of 
$\gamma $. We name such families \textit{Stochastic Procesess with Margins
Estimating of the EXtreme value Index (SPMEEXI's)}. Up to our knowledge, the
first was introduced in Lo \cite{gslotheta} (see also Lo \cite{gslohdarxiv})
as follows

\begin{equation*}
T_{n}(p)=k^{-1}\sum\limits_{h=1}^{p}\sum_{(s_{1}......s_{h})\in \mathcal{P}%
(p,h)}\overset{i_{0}}{\underset{i_{1}=\ell +1}{\sum }}...\sum_{i_{h}=\ell
+1}^{i_{h-1}}i_{h}\prod\limits_{i=i_{1}}^{i_{h}}\frac{\left( \log
X_{n-i+1,n}-\log X_{n-i,n}\right) ^{s_{i}}}{s!},
\end{equation*}

\noindent for $1\leq \ell <k<n,$ $p\geq 1,$\ $i_{0}=k,$ where $\mathcal{P}%
(p,h)$ is the set of all ordered partitions of $p>0$ into positive integers, 
$1\leq h\leq p:$

\begin{equation*}
\mathcal{P}(p,h)=\left\{ (s_{1}...s_{h}),\forall i,\text{ }1\leq i\leq
h,s_{i}>0;s_{1}+...+s_{h}=p\right\} .  \label{alpha7}
\end{equation*}
Further Lo \textit{et al.} \cite{dioplo} and \cite{dioplo90} introduced
continuous and functional forms described in (\ref{tf01}) below. Meanwhile,
without denoting it such that, Segers \cite{segers} and others considered
the Pickands process $\{P_{n}(s),\sqrt{k/n}\leq s\leq 1\},$ with 
\begin{equation*}
P_{n}(s)=\log \frac{X_{n-\left[ k/s\right] ,n}-X_{n-\left[ k/1\right] ,n}}{%
X_{n-\left[ k/s^{2}\right] ,n}-X_{n-\left[ k/s\right] ,n}},\sqrt{k/n}\leq
s\leq 1.
\end{equation*}

\noindent Groeneboom \cite{glw}, proposed a family of kernel estimators,
indexed by kernels. This family is surely a SPMEEXI although the authors did
not consider a stochastic process view in the kernels $K$.

\bigskip

\noindent The main interest of \textit{SPMEEXI}'s is first to have in hands
an infinite class of estimators and especially, as shown in Segers (\cite%
{segers}), to have the possibility to build discrete and continuous
combinations of the margins as new and powerful estimators.

\bigskip

Number of the estimators of the extreme value index are either functions of
consecutive log-spacings $\log X_{n-j+1,n}-\log X_{n-j,n}$ for $1\leq j\leq
k $, $1\leq k\leq n$, or are functions of log-spacings from a threshold $%
\log t $ : $\log X_{n-j+1,n}-\log t$, $1\leq j\leq k$, $1\leq k\leq n$. In
the last case, the threshold is usually taken as $t=X_{n-k,n}$. This simple
remark teases the idea that taking functions of the log-spacings in place of
the simple ones may lead to more general estimators. Dekkers \textit{et al.} 
\cite{deh} successfully experimented to the so-called moment estimator by
using the power functions $h(x)=x^{p}$, $x\in \mathbb{R_{+}}$, $p=1,2.$ Some
available SPMEEX's are functions of these log-spacings as we will see soon.
Here, In this paper, we aim at presenting a more general functional form in
the following 
\begin{equation}
T_{n}(f,s)=\sum_{j=1}^{k}f(j)\left( \log X_{n-j+1,n}-\log X_{n-j,n}\right)
^{s},  \label{tf01}
\end{equation}%
indexed by some classes $\mathcal{F}$ of functions $f:\mathbb{N}^{\ast }=%
\mathbb{N}\backslash \{0\}\longmapsto \mathbb{R}_{+}$, and by $s>0$. We have
two generalizations. First, for $s=1$, we get 
\begin{equation*}
T_{n}(f,1)/k=\sum_{j=1}^{k}f(j)\left( \log X_{n-j+1,n}-\log X_{n-j,n}\right)
/k,
\end{equation*}%
which is the functional generalization of the Diop and Lo statistics \cite%
{dioplo90} for $f(j)=j^{\tau },$ for $0<\tau $ and Deme \textit{et al.} \cite%
{demedioplo}. Secondly, if $f$ is the identity function and $s=1$, we see
that $T_{n}(Identity,1)/k$ is Hill's statistic.\newline

\noindent On the other hand, when utilizing the threshold method, we have,
with the same properties of the parameters, the following statistic process :

\begin{equation}
S_{n}(f,s)=\sum_{j=1}^{k}f(j)\left( \log X_{n-j+1,n}-\log
X_{n-k,n}\right)^{s}, .  \label{tf02}
\end{equation}

\noindent This leads the couple of statistics 
\begin{equation*}
(M_{1,n},M_{1,n})=(S_{n}(\mathbf{1},1)/k,M_{n}(\mathbf{1},2)/k)
\end{equation*}

\noindent where $\mathbf{1}$\ is the constant function $\mathbf{1}(x)=1$\ .
From this couple of statistics Dekkers \textit{et al.} \cite{deh} \ deduced
the following estimator of the extreme value index

\begin{equation*}
D_n=M_{1,n} 1 + (1- (M_{2,n}/M_{1,n}^{2}))^{-1}/2.
\end{equation*}

\bigskip
\noindent Our objective is to show that these two stochastic processes (\ref{tf01}) and (\ref{tf02}) are
SPMEEXI's. In this paper, we focus on the stochastic process $T_{n}(f,s)$
which uses sums of independant random variables. As to $S_{n}(f,s)^{\prime
}s,$ to the contrary, it uses sums of dependent random variables. Its study
will be done in coming up papers.\newline

\subsection{Motivations and scope of the paper}

\noindent As announced, we focus on the stochastic process (\ref{tf01})
here. We have been able to establish its finite-dimension asymptotic
distribution. As already noticed in earlier works in Lo \textit{et al.}
(\cite{dioplo90}, \cite{demedioplo}), the limiting law may be
Gaussian or non-Gaussian. In both cases, statistical tests may be
implemented. In case of non-Gaussian asymptotic limits, the limiting
distribution is represented through an infinite series of standard
exponential random variables. Its law may be approximated through
monte-Carlo methods, as showed in Fall \textit{et al.} \cite{falladja}.%
\newline

\noindent Then we prove that it is a SPMEEXI in the sense of convergence in
probability. Both for asymptotic distribution and convergence in
probability, the used conditions are expressed with respect to an infinite
series of standard exopnential randoms variables and through the auxiliary
functions $a$ and $p$ in the representations of $df$'s in the extreme domain
of attraction that will be recalled in the just next subsection. The
conditions are next notably simplified by supposing that the \textit{df} $%
F $ is differentiable in the neighborhood of its upper endpoint.\newline

\noindent To show how work the results for specific classes of functions $f$%
, we adapt them for $f_{\tau }(j)=j^{\tau }$, $\tau >0$. It is interesting
to see that although we have the existence of the asymptotic laws for any $%
\tau >0$ and $s\geq 1$, we don't have an estimation of $\gamma $ in the
region $\tau <s-1$, when $s>1$.\newline

\noindent One advantage of using SPMEEXI's is that we may consider the best
estimators, in some sense to be precised, among all margins. We show in
Theorem \ref{theo3} that $T_{n}(f_{\tau },s)$ is asymptotically Gaussian for 
$\tau \geq s-1/2$. When we restrict ourselves in that domain, we are able to
establish that the minimum asymptotic variance is reached for $\tau =s$. Then
we construc the best estimator $T_{n}^{(\tau)}=T_{n}(f_{\tau },\tau
) $, that is for $\tau =s$. This is very important since the Hill estimator
is $T_{n}^{(1)}$ itself and, as a consequence, the Hill estimator is an element of a set
of best estimators indexed by $\tau $. In fact, it is the best of all, that is
$T_{n}^{(1)}$ has less asymptotic variance than $T_{n}^{(\tau)}, \tau > 1$.\\

\noindent It will be interesting to found out whether this minimim variance can be
improved for other functional classes.\\

\noindent Even when we have a minimum asymptotic variance estimator, it is not sure that the performance 
is better for finite samples. This is why simulation studies mean reveal a best combination
between bias and asymptotic variance. At finite sample size, the performance of an estimator is measured
both by the bias and the variance and we don't know how the random value of the estimator is far from the exact value.
We will see in the simulation Section \ref{sec4} that the boundary case $\tau=s-1/2$ gives performances similar to the optimal
case.\\

\noindent Before we present the theoritical results and their consequences,
we feel obliged to present a brief reminder of basic univariate extreme
value theory and some related notation on which the statements of the
results will rely on.

\subsection{Basics of Extreme Value Theory}

\label{subsec2}

\noindent Let us make this reminder by continuing the lines of (\ref{dl05})
above. \noindent If (\ref{dl05}) holds, it is said that $F$ is attracted to $%
M$ or $F$ belongs to the domain of attraction of $M$, written $F\in D(M).$
It is well-kwown that the three possible nondegenerate limits in (\ref{dl05}%
), called extreme value \textit{df}, are the following : \newline

\noindent The Gumbel \textit{df} of parameter $\gamma =0,$

\begin{equation}
\Lambda (x)=\exp (-\exp (-x)),\text{ }x\in \mathbb{R},  \label{dl05a}
\end{equation}%
or the Fr\'{e}chet \textit{df} of parameter $\gamma >0,$

\begin{equation}
\phi _{\gamma }(x)=\exp (-x^{-\gamma })\mathbb{I}_{\left[ 0,+\infty \right[
}(x),\text{ }x\in \mathbb{R}\   \label{dl05b}
\end{equation}%
or the Weibull \textit{df} of parameter $\gamma <0$ ,

\begin{equation}
\psi _{\gamma }(x)=\exp (-(x)^{-\gamma })\mathbb{I}_{\left] -\infty ,0\right]
}(x)+(1-1_{\left] -\infty ,0\right] }(x)),\ \ x\in \mathbb{R},\ 
\label{dl05c}
\end{equation}%
where $I_{A}$ denotes the indicator function of the set A. Now put $D(\phi
)=\cup _{\gamma >0}D(\phi _{\gamma }),$ $D(\psi )=\cup _{\gamma >0}D(\psi
_{\gamma }),$ and $\Gamma =D(\phi )\cup D(\psi )\cup D(\Lambda )$.\newline

\bigskip

\noindent In fact the limiting distribution function $M$ is defined by an
equivalence class of the binary relation $\mathcal{R}$ on the set of \textit{%
df} $\mathcal{D}$ on $F$ defined as follows : 
\begin{equation*}
\forall (M_{1},M_{2})\in \mathcal{D}^{2},(M_{1}\text{ }\mathcal{R}\text{ }%
M_{2})\Leftrightarrow \exists (a,b)\in \mathbb{R}_{+}\backslash \{0\}\times 
\mathbb{R},\forall (x\in \mathbb{R}),
\end{equation*}%
\begin{equation*}
M_{2}(x)=M_{1}(ax+b).
\end{equation*}%
One easily checks that if $F^{n}\left( a_{n}x+b_{n}\right) \rightarrow
M_{1}(x),$ then $F^{n}\left( c_{n}x+d_{n}\right) \rightarrow
M_{1}(ax+b)=M_{2}(x)$ whenever 
\begin{equation}
a_{n}/d_{n}\rightarrow a\text{ and }(b_{n}-d_{n})/c_{n}\rightarrow b\text{
as }n\rightarrow \infty .  \label{dl05f}
\end{equation}

\noindent Theses facts allow to parameterize the class of extremal
distribution functions. For this purpose, suppose that (\ref{dl05}) holds
for the three \textit{df}'s given in (\ref{dl05a}), (\ref{dl05b}) and (\ref%
{dl05c}). We may take sequences $(a_{n}>0)_{n\geq 1}$ and $(b_{n})_{n\geq 1}$
such that the limits in (\ref{dl05f}) are $a=\gamma =1/\alpha $ and $b=1$
(in the case of Fr\'{e}chet extremal domain), and $a=-\gamma =-1/\alpha $
and $b=-1$ (in the case of Weibull extremal domain). Finally, one may
interprets $(1+\gamma x)^{-1/\gamma }=exp(-x)$ for $\gamma =0$ (in the case
of Gumbel extremal domain). This leads to the following parameterized
extremal distribution function 
\begin{equation*}
G_{\gamma }(x)=\exp (-(1+\gamma x)^{-1/\gamma }),\text{ }1+\gamma x\geq 0,
\label{dl05d}
\end{equation*}%
called the Generalized Extreme Value (GEV) distribution of parameter $\gamma
\in R$.

\bigskip

\noindent Now we give the usual representations of $df^{\prime }s$ lying in
the extremal domain in terms of the quantile function of $%
G(x)=F(e^{x}),x\geq 1,$ that is $G^{-1}(1-u)=\log F^{-1}(1-u),0\leq u\leq 1.$

\begin{theorem}
\label{theo1} We have :

\begin{enumerate}
\item Karamata's representation (KARARE)

\noindent (a) If $F\in D(\phi _{1/\gamma }),$ $\gamma >0$, then 
\begin{equation}
G^{-1}(1-u)=\log c+\log (1+p(u))-\gamma \log u+(\int_{u}^{1}b(t)t^{-1}dt),%
\text{ }0<u<1,  \label{rep1}
\end{equation}%
where $\sup (\left\vert p(u)\right\vert ,\left\vert b(u)\right\vert
)\rightarrow 0$ as $u\rightarrow 0$ and c is a positive constant and $%
G^{-1}(1-u)=\inf \{x,G(x)\geq u\},$ $0\leq u\leq 1,$ is the generalized
inverse of $G$ with $G^{-1}(0)=G^{-1}(0+)$.\newline
\newline
\noindent (b) If $F\in D(\psi _{1/\gamma }),$ $\gamma >0$, then $%
y_{0}(G)=\sup \{x,$ $G(x)<1\}<+\infty $ and 
\begin{equation}
y_{0}-G^{-1}(1-u)=c(1+p(u))u^{\gamma }\exp \biggr(\int_{u}^{1}b(t)t^{-1}dt%
\biggr),\text{ }0<u<1,  \label{rep2}
\end{equation}%
where $c$, $p(\cdot )$ and $b(\cdot )$ are as in (\ref{rep1})

\item Representation of de Haan (Theorem 2.4.1 in \cite{dehaan}),

\noindent If $G\in D(\Lambda )$, then 
\begin{equation}
G^{-1}(1-u)=d-a(u)+\int_{u}^{1}a(t)t^{-1}dt,\text{ }0<u<1,  \label{rep3a}
\end{equation}%
\noindent where d is a constant and $a(\cdot )$ admits this KARARE : 
\begin{equation}
a(u)=c(1+p(u))\exp (\int_{u}^{1}b(t)t^{-1}dt),\text{ }0<u<1,  \label{rep3b}
\end{equation}%
\noindent $c$, $p(\cdot )$ anf $b(\cdot )$ being defined as in (\ref{rep1}).
We warn the reader to not confuse this function $a(.)$ with the function $%
a_{n}(.,.)$ which will be defined later.
\end{enumerate}
\end{theorem}

\bigskip

\noindent Finally, we shall also use the uniform representation of $%
Y_{1}=\log X_{1},Y_{2}=\log X_{2},...$ by $%
G^{-1}(1-U_{1}),G^{-1}(1-U_{2}),...$ where $U_{1},U_{2},...$ are independent
and uniform random variables on $(0,1)$ and where $G$ is the $df$ of $Y$, in
the sense of equality in distribution (denoted by $=_{d})$%
\begin{equation*}
\left\{ Y_{j},j\geq 1\}=_{d}\{G^{-1}(1-U_{j}),j\geq 1\right\} ,
\end{equation*}%
and hence 
\begin{equation}
\{\left\{ Y_{1,n},Y_{2,n},...Y_{n,n}\right\} ,n\geq 1\}  \label{repre}
\end{equation}

\begin{equation*}
=_{d}\left\{
\{G^{-1}(1-U_{n,n}),G^{-1}(1-U_{n-1,n}),...,G^{-1}(1-U_{1,n})\},n\geq
1\right\} .
\end{equation*}

\bigskip

\noindent In connexion with this, we shall use the following Malmquist
representation (see \cite{shwell}, p. 336) :

\begin{equation*}
\{\log (\frac{U_{j+1,n}}{U_{j,n}})^{j},j=1,...,n\}=_{d}\{E_{1,n},...,E_{n,n}%
\},  \label{malm}
\end{equation*}%
where $E_{1,n},...,E_{n,n}$ is an array of independent standard exponential
random variables. We write $E_{i}$ instead of $E_{i,n}$ for simplicity sake.
Some conditions will be expressed in terms of these exponential random
variables. \noindent We are now in position to state our first results for
finite distribution asymptotic normality.

\section{Our results} \label{sec2}

We need the following conditions. First define for $n\geq 1,$ $f$ and $s$
fixed,

\begin{equation*}
B_{n}(f,s)=\max \{f(j)j^{-s}/\sigma _{n}(f,s),1\leq j\leq k\},
\end{equation*}

\begin{equation*}
a_{n}\left( f,s\right) =\Gamma(s+1) \sum_{j=1}^{k}f\left( j\right) j^{-s}
\end{equation*}

\noindent and

\begin{equation*}
\sigma _{n}^{2}\left( f,s\right) =\sum_{j=1}^{k}f^{2}\left( j\right) j^{-2s}.
\end{equation*}%
We will use the two main conditions of $f$ and $s$ fixed : 
\begin{equation}
\sum_{j=1}^{\infty }f(j)^{2}j^{-2s}<\infty  \tag{K1}
\end{equation}%
and%
\begin{equation}
\sum_{j=1}^{\infty }f(j)^{2}j^{-2s}=+\infty \text{ and }B_{n}(f,s)%
\rightarrow 0\text{ as }n\rightarrow \infty .  \tag{K2}
\end{equation}

\noindent Further, any \textit{df} in $D(G_{\gamma })$ is associated with a
couple of functions $(p,b)$ as given in the representations (\ref{rep1}), (%
\ref{rep2}) and (\ref{rep3b}). Define then the following notation for $%
\lambda >1$,

\begin{equation*}
b_{n}\left( \lambda \right) =\sup \{\left\vert b(t)\right\vert ,0\leq t\leq
\lambda k/n\}
\end{equation*}

\noindent and

\begin{equation*}
\ p_{n}\left( \lambda \right) =\sup \{\left\vert p(t)\right\vert ,0\leq
t\leq \lambda k/n\}
\end{equation*}%
We will require below that, for some $\lambda >1,$%
\begin{equation}
b_{n}\left( \lambda \right) \log k\rightarrow 0, \text{ as } n \rightarrow
+\infty.  \tag{CR1}
\end{equation}

\noindent From now, all the limits \ below \ are meant as $n\rightarrow
\infty $ unless the contrary \ is specified.

\noindent Here are our fundamental results. First, we have marginal estimations
of the extreme value index as expected. The conditions of the results are
given in very general forms that allow further, specific hypotheses as
particular cases. As well, although we focus here on finite-distribution
limits, the conditions are stated in a way that will permit to handle
uniform studies further.

\begin{theorem}
\label{theo2} Let $F\in D(G_{\gamma })$, $0\leq \gamma <+\infty $.\newline

\noindent (A) Case $0<\gamma <+\infty $.\newline

\noindent 1) Let $(K1)$ \ hold. If $a_{n}(f,s)\rightarrow \infty $ and for
an arbitrary $\lambda >1$

\ 
\begin{equation}
a_{n}^{-1}\left( f,s\right) \left[ \sum_{j=1}^{k}f\left( j\right) s\left\{ 
\frac{\gamma }{j}E_{j}+\left( p_{n}(\lambda )+\frac{E_{j}}{j}b_{n}(\lambda
)\right) \right\} ^{s-1}\right.  \tag{H0a}
\end{equation}

\begin{equation*}
\times \left. \left( p_{n}(\lambda )+\frac{E_{j}}{j}b_{n}(\lambda )\right) %
\right] \rightarrow _{\mathbb{P}}0,
\end{equation*}

\noindent then

\begin{equation*}
(T_{n}(f,s)/a_{n}\left( f,s\right) )^{1/s}\rightarrow _{\mathbb{P}}\gamma ,
\end{equation*}

\noindent where $\rightarrow _{\mathbb{P}}$ stands for convergence in
probability.

\bigskip \bigskip

\noindent 2) Let $(K2)$ hold. If \ $a_{n}^{-1}(f,s)\sigma
_{n}^{-1}(f,s)\rightarrow 0$\ and for an arbitrary $\lambda >1$

\begin{equation}
\sigma _{n}^{-1}(f,s)\left[ \sum_{j=1}^{k}f\left( j\right) s\left\{ \frac{%
\gamma }{j}E_{j}+\left( p_{n}(\lambda )+\frac{E_{j}}{j}b_{n}(\lambda
)\right) \right\} ^{s-1} \right.  \tag{H1a}
\end{equation}

\begin{equation*}
\times \left. \left( p_{n}(\lambda )+\frac{E_{j}}{j}b_{n}(\lambda )\right) %
\right] \rightarrow _{\mathbb{P}}0,
\end{equation*}

\noindent then 
\begin{equation*}
(T_{n}(f,s)/a_{n}\left( f,s\right) )^{1/s}\rightarrow _{\mathbb{P}}\gamma .
\end{equation*}

\bigskip

\noindent (B) case $\gamma =0$.\newline

\noindent 1) Let $(K1)$ \ hold. If $a_{n}\left( f,s\right) \rightarrow
+\infty $ and for an arbitrary $\lambda >1$,

\begin{equation}
a_{n}^{-1}\left( f,s\right) \left[ \sum_{j=1}^{k}f\left( j\right) s\left\{
j^{-1}E_{j}+\left( p_{n}(\lambda )+\frac{E_{j}}{j}p_{n}(\lambda )\vee
b_{n}(\lambda )\log k\right) \right\} ^{s-1}\right.  \tag{H0b}
\end{equation}

\begin{equation*}
\times \left. \left( p_{n}(\lambda )+\frac{E_{j}}{j}p_{n}(\lambda )\vee
b_{n}(\lambda )\log k\right) \right] \rightarrow _{\mathbb{P}}0,
\end{equation*}

then 
\begin{equation*}
\left( \frac{T_{n}\left( f,s\right) }{a_{n}\left( f,s\right) }\right)
^{1/s}/a(k/n)\rightarrow _{\mathbb{P}}1.
\end{equation*}

\bigskip \bigskip 2) Let $(K2)$ \ hold. If $a_{n}^{-1}(f,s)\sigma
_{n}^{-1}(f,s)\rightarrow 0$ and for an arbitrary $\lambda >1$, and

\begin{equation}
\sigma _{n}^{-1}\left( f,s\right) \left[ \sum_{j=1}^{k}f\left( j\right)
s\left\{ j^{-1}E_{j}+\left( p_{n}(\lambda )+\frac{E_{j}}{j}p_{n}(\lambda
)\vee b_{n}(\lambda )\log k\right) \right\} ^{s-1}\right.  \tag{H1b}
\end{equation}

\begin{equation*}
\times \left. \left( p_{n}(\lambda )+\frac{E_{j}}{j}p_{n}(\lambda )\vee
b_{n}(\lambda )\log k\right) \right] \rightarrow _{\mathbb{P}}0,
\end{equation*}

then 
\begin{equation*}
\left( \frac{T_{n}\left( f,s\right) }{a_{n}\left( f,s\right) }\right)
^{1/s}/a(k/n)\rightarrow _{\mathbb{P}}1.
\end{equation*}
\end{theorem}

\bigskip \bigskip \bigskip

\begin{theorem}
\label{theo3}

Let $F\in D(G_{\gamma })$, $0\leq \gamma <+\infty $.\newline

\noindent (A) Case $0<\gamma <+\infty $.\newline

\noindent 1) If $(K1)$ \ and 
\begin{equation}
\sum_{j=1}^{k}f\left( j\right) s\left\{ \frac{\gamma }{j}E_{j}+\left(
p_{n}(\lambda )+\frac{E_{j}}{j}b_{n}(\lambda )\right) \right\} ^{s-1}\left(
p_{n}(\lambda )+\frac{E_{j}}{j}b_{n}(\lambda )\right) \rightarrow _{\mathbb{P%
}}0,  \tag{H2a}
\end{equation}

\noindent then 
\begin{equation*}
T_{n}(f,s)-\gamma ^{s}a_{n}\left( f,s\right) \rightarrow \gamma ^{s}\left\{
\Gamma \left( 2s+1\right) -\Gamma \left( s+1\right) ^{2}\right\} ^{1/2}%
\mathcal{L}(f,s),
\end{equation*}

\noindent where 
\begin{equation*}
\mathcal{L}(f,s)=\sum_{j=1}^{\infty }f\left( j\right) j^{-s}F_{j}^{(s)},
\end{equation*}%
and the $F_{j}^{s}$'s are independent and centred random variables with
variance one.

\bigskip \noindent 2) If $(K2)$ \ and $(H2a)$ hold for an arbitrary $\lambda
>1$, then 
\begin{equation*}
\sigma _{n}^{-1}(f,s)\left( T_{n}(f,s)-\gamma ^{s}a_{n}\left( f,s\right)
\right) \rightarrow \mathcal{N}(0,\gamma ^{2s}\left\{ \Gamma \left(
2s+1\right) -\Gamma \left( s+1\right) ^{2}\right\} ).
\end{equation*}

\bigskip \bigskip \noindent (B) case $\gamma =0$.\newline

\noindent 1) If $(K1)$ \ and for an arbitrary $\lambda >1$, 
\begin{equation}
\left[ \sum_{j=1}^{k}f\left( j\right) s\left\{ j^{-1}E_{j}+\left(
p_{n}(\lambda )+\frac{E_{j}}{j}p_{n}(\lambda )\vee b_{n}(\lambda )\log
k\right) \right\} ^{s-1}\right.  \tag{H2b}
\end{equation}

\begin{equation*}
\times \left. \left( p_{n}(\lambda )+\frac{E_{j}}{j}p_{n}(\lambda )\vee
b_{n}(\lambda )\log k\right) \right] \rightarrow _{\mathbb{P}}0,
\end{equation*}

then 
\begin{equation*}
\frac{T_{n}\left( f,s\right) }{a^{s}\left( \frac{k}{n}\right) }-a_{n}\left(
f,s\right) \rightarrow \left\{ \Gamma \left( 2s+1\right) -\Gamma \left(
s+1\right) ^{2}\right\} ^{1/2}\mathcal{L}(f,s)),
\end{equation*}%
where 
\begin{equation*}
\mathcal{L}(f,s)=\sum_{j=1}^{\infty }f\left( j\right) j^{-s}F_{j}^{(s)}.
\end{equation*}

2) If $(K2)$ \ and $(H2b)$ hold for an arbitrary $\lambda >1$, then 
\begin{equation*}
\sigma _{n}^{-1}(f,s)\left[ \frac{T_{n}\left( f,s\right) }{a^{s}\left(
k/n\right) }-a_{n}\left( f,s\right) \right] \rightarrow \mathcal{N}\left(
0,\left\{ \Gamma \left( 2s+1\right) -\Gamma \left( s+1\right) ^{2}\right\}
\right) .
\end{equation*}
\end{theorem}

\subsection{Remarks and applications.}

\subsubsection{General remarks on the conditions}

The conditions $(H0a)$, $(H0b)$, $(H1a)$, $(H1b)$, $(H2a)$ and $(H2b)$ hold
if we show that the expectations of the \textit{rv}'s of their right members
tend to zero, for an arbitrary $\lambda >1$, simply by the use of Markov's
inequality. These expectations include intergrals $I(a,b,s)=\int_{0}^{\infty
}(a+bx)^{s}e^{-x}dx$ and $J(a,b,c,s)=\int_{0}^{\infty
}(a+cx)^{s}(a+bx)e^{-x}dx$ for real numbers $a$, $b$, $c$ and $s\geq 1$
computed in $(4.1)$\ and $(4.2)$\ of Section (\ref{sectiontool}), \textbf{\textit{for integer values of $s$}}, as 
\begin{equation*}
I(a,b,s)=\int_{0}^{\infty
}(a+bx)^{s}e^{-x}dx=s!\sum_{h=0}^{s}b^{h}a^{s-h}/(s-h)!.
\end{equation*}%
and 
\begin{eqnarray*}
J(a,b,c,s)
&=&s!a\sum_{h=0}^{s-1}c^{h}a^{s-h}/(s-h)!+s!b%
\sum_{h=0}^{s-1}c^{h}I(a,c,s-h)/(s-h)! \\
&+&s!c^{s}I(a,b,1).
\end{eqnarray*}%
Then the conditions $(H0a),(H0b),(H2a),(H2b),(H1a)$ and $(H1b)$ respectively
hold when hold these ones

\begin{tabular}{ll}
(HE0a) & $a_{n}^{-1}(f,s)\sum_{j=1}^{k}f(j)I_{n}(1,j,s)\rightarrow 0$; \\ 
(HE0b) & $a_{n}^{-1}(f,s)\sum_{j=1}^{k}f(j)I_{n}(2,j,s)\rightarrow 0$; \\ 
(HE1a) & $\sigma_{n}^{-1}(f,s)\sum_{j=1}^{k}f(j)I_{n}(1,j,s)\rightarrow 0$;
\\ 
(HE1b) & $\sigma_{n}^{-1}(f,s)\sum_{j=1}^{k}f(j)I_{n}(2,j,)\rightarrow 0$;
\\ 
(HE2a) & $\sum_{j=1}^{k}f(j)I_{n}(1,j,s)\rightarrow 0$; \\ 
(HE2b) & $\sum_{j=1}^{k}f(j)I_{n}(2,j,s)\rightarrow 0$; \\ 
& 
\end{tabular}

\noindent with%
\begin{equation*}
I_{n}(1,j,s)=sJ(p_{n}(\lambda ),b_{n}(\lambda )/j,(\gamma +b_{n}(\lambda
))/j,s-1)
\end{equation*}%
and%
\begin{equation*}
I_{n}(2,j,s)=sJ(p_{n}(\lambda ),\left( p_{n}(\lambda )\vee b_{n}(\lambda
)\log k\right) /j,(1+(p_{n}(\lambda )\vee b_{n}(\lambda )\log k))/j,s-1)
\end{equation*}

\subsection{Weakening the conditions for $s$ interger.}

When the distribution function $G$ admits an ultimate derivative at $%
x_{0}(G)=\sup \{x,G(x)<1\},$ and this is the case for the usual \textit{df}%
's, one may take $p(u)=0$, as pointed out in \cite{fall-lo}. In that case,
the conditions $(HE0x),(HE1x)$ and ($HE2x),$ for $x=a$ or $x=b,$ are much
simpler. We then have $I_{n}(1,j,s)=ss!b_{n}(\lambda )(\gamma +b_{n}(\lambda
))^{s-1}j^{-s}$ and $I_{n}(2,j,s)=ss!(b_{n}(\lambda )\log k)(\gamma
+b_{n}(\lambda )\log k)^{s-1}j^{-s}.$ We get these simpler conditions :

\begin{tabular}{ll}
(HE0a) & $a_{n}^{-1}(f,s)\left( ss!b_{n}(\lambda )(\gamma +b_{n}(\lambda
))^{s-1}\sum_{j=1}^{k}f(j)j^{-s}\right) \rightarrow 0$; \\ 
(HE0b) & $a_{n}^{-1}(f,s)\left( ss!(b_{n}(\lambda )\log
k)^{s-1}\sum_{j=1}^{k}f(j)j^{-s)}\right) \rightarrow 0$; \\ 
(HE1a) & $\text{ }\sigma _{n}^{-1}(f,s)\left( ss!b_{n}(\lambda )(\gamma
+b_{n}(\lambda ))^{s-1}\sum_{j=1}^{k}f(j)j^{-s}\right) \rightarrow 0$; \\ 
(HE1b) & $\text{ }\sigma _{n}^{-1}(f,s)\left( ss!(b_{n}(\lambda )\log
k)^{s-1}\sum_{j=1}^{k}f(j)j^{-s)}\right) \rightarrow 0$; \\ 
(HE2a) & $\left( ss!b_{n}(\lambda )(\gamma +b_{n}(\lambda
))^{s-1}\sum_{j=1}^{k}f(j)j^{-s)}\right) \rightarrow 0$; \\ 
(HE2b) & $\left( ss!(b_{n}(\lambda )\log k)(\gamma +b_{n}(\lambda )\log
k)^{s-1}\sum_{j=1}^{k}f(j)j^{-s)}\right) \rightarrow 0$.%
\end{tabular}

\bigskip

\noindent It is interesting to remark that all these conditions automaticaly
hold whenever $b_{n}(\lambda )\rightarrow 0,$ and/or $(CR1)$ holds. Indeed, we
remark, by the Cauchy-Scharwz's inequality, that :%
\begin{equation*}
\sigma _{n}^{-1}(f,s)\sum_{j=1}^{k}f(j)j^{-s}\leq \sigma
_{n}^{-1}(f,s)\left( \sum_{j=1}^{k}f^{2}(j)j^{-2s}\right) ^{1/2}=1.
\end{equation*}%
Then for $\gamma >0,$ the corresponding conditions always hold since $%
b_{n}(\lambda )\rightarrow 0$ and for $\gamma =0,$ the corresponding
conditions hold with $(CR1)$. This surely leads to powerfull results. It
also happens that for the usual cases, we know the values of $b_{n}(\lambda
) $, based on $b(u)=(G^{-1}(1-u))^{\prime }+\gamma $ for $\gamma >0$ and $%
b(u)=us^{\prime }(u)/s(u)$ with $s(u)=u(G^{-1}(1-u))^{\prime }$ (see for
instance \cite{fall-lo} or \cite{segers})

\subsection{The special case of Diop-Lo}

\label{sbdl}

Now it is time to see how the preceeding results work for the particular
case the functions class $f_{\tau }(j)=j^{\tau },\tau >0.$ This special
study should \ be a model of \ how to apply the results for other specific
classes. Here, we will replace $f$ by $\tau $ in all the notation meaning
that $f=f_{\tau }$. We summarize the holding conditions depending on $\tau
>0 $ and $s\geq 1$, in the following table

\begin{table}[tbp]
\begin{tabular}{llll}
\hline\hline
(I) & (II) & (III) & (IV) \\ \hline\hline
$\tau <s-1$ & $s-1\leq \tau <s-1/2$ & $\tau =s-1/2$ & $\tau >s-1/2$ \\ 
$(K1)$ & $(K1)$ & $(K2)$ & $(K2)$ \\ \hline\hline
$a_{n}$ bounded & $a_{n}\rightarrow \infty $ & $a_{n}\sim 2s! k^{1/2}$ & $%
a_{n}\sim s!\frac{k^{\tau-s+1}}{\tau -s+1}$ \\ 
&  &  &  \\ 
$\sigma _{n}$ bounded & $\sigma _{n}$ bounded & $\sigma _{n}\sim (\log
k)^{1/2}$ & $\sigma _{n}\sim \frac{k^{(\tau -s+1/2)}}{\sqrt{2(\tau -s)+1}}$
\\ \hline\hline
\end{tabular}%
\caption{Checking the conditions for the Diop-Lo class}
\label{tabdioplo}
\end{table}
\bigskip \noindent We may see the details as follows. First $\sum
f(j)^{2}j^{-2s}=\sum j^{-2(s-\tau )}$ is finite if and only if $2(s-\tau
)>1. $ This gives the cases (I) and (II). For (III) in Table \ref{tabdioplo}%
, we have 
\begin{equation*}
\sigma _{n}^{2}(\tau ,s)=\sum_{j=1}^{k}j^{-2(s-\tau
)}=\sum_{j=1}^{k}j^{-1}\sim (\log k),
\end{equation*}%
by (\ref{integ1}) in Section (\ref{sectiontool}). Since for $1\leq j\leq
k,f(j)j^{-s}/\sigma _{n}(\tau ,s)=j^{-1/2}/\sigma _{n}(\tau ,s)\leq 1/\sigma
_{n}(\tau ,s),$ we have $B_{n}(\tau ,s)\leq \sigma _{n}^{-1}(\tau
,s)\rightarrow 0$ and then $(K2)$ holds. For (IV), 
\begin{equation*}
\sigma _{n}^{2}(\tau ,s)=\sum_{j=1}^{k}j^{-2(s-\tau )}\sim
k^{(2(\tau-s)+1))}/(2(\tau-s)+1)),
\end{equation*}%
by (\ref{integ3}). Next $1\leq j\leq k,$ $f(j)j^{-s}/\sigma _{n}(\tau
,s)=j^{(\tau -s)}/\sigma _{n}(\tau ,s)=C_{0}(j/k)^{(\tau -s)}k^{-1/2}.$
Since $(\tau -s)>0,B_{n}(\tau ,s)\leq C_{0}k^{-1/2}\rightarrow 0.$ Then $%
(K2) $ also holds. The lines above also explain the fourth row of the table.
The third row is immediate since $a_{n}=\sum_{j=1}^{k}j^{-(s-\tau
)}\rightarrow \infty $ for $(s-\tau )\leq 1$ and remains bounded for $%
(s-\tau )>1.$ 

\bigskip \noindent It is worth mentioning that the case $\tau <s-1$ is not
possible for $s=1$. This unveils a new case comparatively with former
studies of Deme \textit{et al.} \cite{demedioplo} for $s=1$. 

\bigskip \noindent Now, based on these facts, we are able to get the following estimations (results $(CR1)$ for $\gamma=0$) :

\begin{enumerate}
\item For $s-1\leq \tau <s-1/2,$ $a_{n}\rightarrow \infty $. $\ $Hence we
have the estimation 
\begin{equation*}
\left( \frac{T_{n}\left( \tau ,s\right) }{a_{n}\left( \tau ,s\right) }%
\right) ^{1/s}/a(k/n)\rightarrow _{\mathbb{P}}1.
\end{equation*}%
For $\gamma >0,$ $a(k/n)\rightarrow \gamma .$ For $\gamma =0,$ $\left( \frac{%
T_{n}\left( \tau ,s\right) }{a_{n}\left( \tau ,s\right) }\right)
^{1/s}\rightarrow \gamma =0$ at the rate of $a_{n}(k/n).$

\item $0<\tau <s-1.$ We do not have an estimation of $\gamma .$
\end{enumerate}

\bigskip

\noindent For testing the hypothesis $F\in D(G_{\gamma }),$ $\gamma \geq 0,$
we derive the following laws by the delta-method under $(CR1)$, especially
for $\gamma =0$.

\bigskip \noindent Let $s-1\leq \tau <s-1/2.$ For $\gamma >0,$%
\begin{equation*}
a_{n}(\tau ,s)\left\{ \left( \frac{T_{n}\left( \tau ,s\right) }{a_{n}\left(
\tau ,s\right) }\right) -\gamma ^{s}\right\}
\end{equation*}

\begin{equation}
\rightarrow \gamma ^{s}\left\{ \Gamma \left( 2s+1\right) -\Gamma \left(
s+1\right) ^{2}\right\} ^{1/2}\mathcal{L}(\tau ,s).  \label{nongauss}
\end{equation}

\noindent For $\gamma =0$%
\begin{equation*}
a_{n}(\tau ,s)\left\{ a(k/n)^{-s}\left( \frac{T_{n}\left( \tau ,s\right) }{%
a_{n}\left( \tau ,s\right) }\right) -1\right\} \rightarrow \left\{ \Gamma
\left( 2s+1\right) -\Gamma \left( s+1\right) ^{2}\right\} ^{1/2}\mathcal{L}%
(\tau ,s).
\end{equation*}

\bigskip \noindent Let $\tau \geq s-1/2.$ In this case $a_{n}(\tau ,s)\sigma
_{n}^{-1}(\tau ,s)\rightarrow +\infty .$ This enables the delta-method
application to the limit in Theorem \ref{theo3}, case (A-2), that is 
\begin{equation}
\frac{a_{n}(\tau ,s)}{\sigma _{n}(\tau ,s)}\left\{ \left( \frac{T_{n}\left(
\tau ,s\right) }{a_{n}\left( \tau ,s\right) }\right) -\gamma ^{s}\right\}
\rightarrow \mathcal{N}(0,\gamma ^{2s}\left\{ \Gamma \left( 2s+1\right)
-\Gamma \left( s+1\right) ^{2}\right\} ).  \label{pss}
\end{equation}
We derive%
\begin{equation*}
\frac{a_{n}(\tau ,s)}{\sigma _{n}(\tau ,s)}\left\{ \left( \frac{T_{n}\left(
\tau ,s\right) }{a_{n}\left( \tau ,s\right) }\right) ^{1/s}-\gamma \right\}
\rightarrow \mathcal{N}(0,(\gamma /s)\left\{ \Gamma \left( 2s+1\right)
-\Gamma \left( s+1\right) ^{2}\right\} )
\end{equation*}

\noindent For $\gamma =0$%
\begin{equation*}
\frac{a_{n}(\tau ,s)}{\sigma _{n}(\tau ,s)}\left\{ a(k/n)^{-s}\left( \frac{%
T_{n}\left( \tau ,s\right) }{a_{n}\left( \tau ,s\right) }\right) -1\right\}
\rightarrow \mathcal{N}(0,\left\{ \Gamma \left( 2s+1\right) -\Gamma \left(
s+1\right) ^{2}\right\} ).
\end{equation*}

\bigskip 
\noindent For the new case $\tau <s-1,$ we have for $\gamma >0,$%
\begin{equation*}
T_{n}(\tau ,s)-A(\tau ,s)\gamma ^{s}\rightarrow \gamma ^{s}\left\{ \Gamma
\left( 2s+1\right) -\Gamma \left( s+1\right) ^{2}\right\} ^{1/2}\mathcal{L}%
(\tau ,s),
\end{equation*}

\noindent where $A(\tau ,s)=\sum_{j=1}^{\infty }j^{-(s-\tau )}<\infty ,$ for 
$\gamma =0,$%
\begin{equation*}
T_{n}(\tau ,s)/a^{s}(k/n)-A(\tau ,s)\rightarrow \left\{ \Gamma \left(
2s+1\right) -\Gamma \left( s+1\right) ^{2}\right\} ^{1/2}\mathcal{L}(\tau
,s).
\end{equation*}

\bigskip

\noindent These two limiting laws also allow statistical tests based on
Monte-Carlo methods as in \cite{demedioplo}.

\subsection{Best performance estimators}

\bigskip In pratical situations, we have to select a particular function $f$
from a particular class $\mathcal{F}$ of \ function $f$. A natural question
is to select a couple $(f,s)$ for which the estimator is the best in some
sense. Here we consider the class \ of Diop-Lo, $f(j)=j^{\tau }$ and \ we
are interested in finding the best \ performance of the estimator $\left( 
\frac{T_{n}(\tau ,s)}{a_{n}(\tau ,s)}\right) ^{1/s}$ of $\gamma >0$. We
place ourselves in the normality domain, that is $\tau >s-1/2$.
Straigthforward computations from (\ref{pss}) yield

\begin{equation*}
V_{n}(\tau ,s)\left[ \left( \frac{T_{n}(\tau ,s)}{a_{n}(\tau ,s)}\right)
^{1/s}-\gamma \right] \leadsto \boldsymbol{N}(0,1)
\end{equation*}

\noindent with

\begin{equation*}
\ V_{n}(\tau ,s)=\frac{a_{n}(\tau ,s)}{\sigma _{n}(\tau ,s)}\times \frac{%
s\gamma ^{-1}}{\sqrt{\Gamma (2s+1)-\Gamma (s+1)^{2}}}.
\end{equation*}

\bigskip 
\noindent So, finding the best performance is achieved for minimum value for
the asymptotic variance $V_{n}(\tau ,s)^{-2}$. We then have to find the
greatest value of variance $V_{n}(\tau ,s)$. But maximizing this function
both in $s$ and $\tau$ might be tricky. However, for \ a fixed $s \geq 1$,
we may find that the maximum value of $V_{n}(\tau ,s)$ for $\tau \in [s-1/2,
+\infty [$. First, we have to isolate the boundary point $\tau=s-1/2$. We
prove in Subsection \ref{minimizationV} below that the maximum value of $%
V_{n}(s,\tau)$ is reached when $\tau=s$. Using the formulae in Table \ref%
{tabdioplo}, we see that, for $\tau \geq s-1/2$, we have

\begin{equation}
\ V_{n}(\tau ,\tau )=\tau !\sqrt{(}k)\times \frac{s\gamma ^{-1}}{\sqrt{%
\Gamma (2\tau +1)-\Gamma (\tau +1)^{2}}}  \label{minGen}
\end{equation}

\noindent and for $\tau=s-1/2$, we get

\begin{equation}
\ V_{n}(\tau ,\tau +1/2)=2s!\sqrt{(}k/\log k)\times \frac{s\gamma ^{-1}}{%
\sqrt{\Gamma (2(\tau +1))-\Gamma (\tau +3/2)^{2}}}.  \label{minFront}
\end{equation}

\noindent We get as best estimator with least asymptotic variance 
\begin{equation*}
T_{n}(\tau)^{(\tau)} =\left( \frac{T_{n}(\tau ,\tau)}{a_{n}(\tau ,\tau)}%
\right) ^{1/s}
\end{equation*}

\noindent for the normality zone $\tau \geq s-1/2$. Its asymptotic
variance (\ref{minGen}) increases when $\tau $ decreases. This means that
the Hill estimator is the best with respect to this sense. \noindent Now, let us move to the non-Gaussian zone, that is $0\leq s-1\leq \tau <s-1/2$%
, corresponding to the column II in the Table \ref{tabdioplo}. We may easily
derive from (\ref{nongauss}) that the asymptotic variance is of equivalent to 
\begin{equation*}
\gamma \frac{\Gamma (2s+1)-\Gamma (s+1)^{2}}{s(\tau -s+1)},
\end{equation*}

\noindent which is still dominated by $V_{n}(\tau ,\tau )^{-1}$. To sum up,
we say that the Hill estimator has best asymptotic variance for all margins.%
\newline

\noindent However for finite sample, we do not know how far the centered and
normalized statistic is from the limiting Gaussian variable or the
non-Gaussan limiting random. Here we are obliged to back on simulation
studies. Let us consider 
\begin{equation}
T_{n}^{(\tau )}=\left( \frac{T_{n}(\tau ,\tau )}{a_{n}(\tau ,\tau )}\right)
^{1/\tau }  \label{DH}
\end{equation}%
and 
\begin{equation*}
T_{n}^{(\tau +1/2)}=\left( \frac{T_{n}(\tau ,\tau +1/2)}{a_{n}(\tau ,\tau +1/2 )}%
\right) ^{1/(\tau +1/2)} \label{ODH}
\end{equation*}

\noindent We get that these two estimators generally  behave better than the
Hill's and the Dekkers \textit{et al.}'s ones. At least, they have equivalent performances. But absolutely, they seem to be more stable in a sense to precised  later. This must result in lesser biases that constitute a compensation of their
poorer performance regarding the asymptotic variance point of view. A full
report of simulation studies are given in Section \ref{sec4}.

\bigskip

\noindent But we should keep in mind that the results presented here, go far beyond
the Diop-Lo family for which the Hill's estimator demonstates to be the
least asymptotic variance estimator.

\bigskip

\noindent Further, researches will be conducted on other functions families in order
to possibly find out estimators with asymptotic variances better that $1/V_{n}(1,1).$

\subsection{ PROOFS} \label{sec3}

We will prove both theorems together. For both cases $\gamma >0$ and $\gamma
=0,$ we will arrive at the final statement based on the hypotheses $(K1)$ or 
$(K2)$, $(H1a)$ or $(H1a),(H2a)$ in the case $\gamma >0$ (A) , and on $(K1)$
or $(K2)$, $(H0b)$ or $(H1b),(H2b)$ in the case $\gamma =0$ $(B).$ In each
case, an analysis will give the corresponding parts (1) and (2) for the two
theorems. We begin with the A case.\newline

\bigskip

\noindent Case (A) : Here, $F\in D(G_{\gamma })$ with $0<\gamma <+\infty $.
\bigskip By using (\ref{repre}), we have 
\begin{equation*}
T_{n}(f,s)=\sum%
\limits_{j=1}^{k}f(j)(G^{-1}(1-U_{j,n})-G^{-1}(1-U_{j+1,n}))^{s}
\end{equation*}

\noindent By (\ref{rep1}), we also have 
\begin{equation*}
G^{-1}(1-u)=\log c+\log (1+p(u))-\gamma \log
u+\int\limits_{u}^{1}b(t)t^{-1}dt,0<u<1.
\end{equation*}

\noindent For $1\leq j\leq k,$

\begin{eqnarray*}
G^{-1}(1-U_{j,n})-G^{-1}(1-U_{j+1,n}) &=&\log \left[ \frac{1+p(U_{j,n})}{%
1+p(U_{j+1,n})}\right] \\
+\gamma \log \left( \frac{U_{j+1,n}}{U_{j,n}}\right)
&+&\int\limits_{U_{j,n}}^{U_{j+1,n}}b(t)t^{-1}dt.
\end{eqnarray*}

\noindent Put $p_{n}=\sup \{|p(t)|,0\leq t\leq U_{k+1,n}\}$ and $b_{n}=\sup
\{|b(t)|,0\leq t\leq U_{k+1,n}\}$. Both $b_{n}$ and $p_{n}$ tend to zero in
probability as $n\rightarrow +\infty$, since $U_{k+1,n}\rightarrow 0$ when $%
(n,k/n)\rightarrow (+\infty ,0)$. We then get

\begin{equation*}
T_{n}(f,s)=\sum\limits_{j=1}^{k}f(j)\left\{ O_{\mathbb{P}}(p_{n})+\frac{%
\gamma }{j}E_{j}+\frac{E_{j}}{j}O_{\mathbb{P}}(b_{n})\right\} ^{s}
\end{equation*}%
Put

\begin{equation*}
A_{n,j}=\left( O_{\mathbb{P}}(p_{n})+\frac{\gamma }{j}E_{j}+\frac{E_{j}}{j}%
O_{\mathbb{P}}(b_{n})\right) ^{s}.
\end{equation*}

\noindent By the mean value Theorem, we have for $s\geq 1$

\bigskip

\begin{eqnarray*}
A_{n,j}-\left( \frac{\gamma }{j}E_{j}\right) ^{s} &=&s\left\{ \frac{\gamma }{%
j}E_{j}+\theta _{n,j}\left( O_{\mathbb{P}}(p_{n})+\frac{E_{j}}{j}O_{\mathbb{P%
}}(b_{n})\right) \right\} ^{s-1} \\
&\times &\left( O_{\mathbb{P}}(p_{n})+\frac{E_{j}}{j}O_{\mathbb{P}%
}(b_{n})\right)
\end{eqnarray*}

\noindent where for $\left\vert \theta _{n,j}\right\vert \leq 1.$ Put%
\begin{equation}
\zeta _{n,j}(s)=s\left\{ \frac{\gamma }{j}E_{j}+\theta _{n,j}\left( O_{%
\mathbb{P}}(p_{n})+\frac{E_{j}}{j}O_{\mathbb{P}}(b_{n})\right) \right\}
^{s-1}\left( O_{\mathbb{P}}(p_{n})+\frac{E_{j}}{j}O_{\mathbb{P}%
}(b_{n})\right)  \label{eq01}
\end{equation}%
We have

\begin{equation*}
T_{n}(f,s)=\gamma
^{s}\sum\limits_{j=1}^{k}f(j)j^{-s}E_{j}^{s}+\sum\limits_{j=1}^{k}f(j)\zeta
_{n,j}(s).
\end{equation*}

\noindent Recall that $\mathbb{E}\left( E_{j}^{s}\right) =\Gamma \left(
s+1\right) $ and $\mathbb{V}(E_{j}^{s})=\Gamma \left( 2s+1\right) -\Gamma
\left( s+1\right) ^{2}$ and denote

\begin{equation*}
V_{n}\left( f,s\right) =\sum_{j=1}^{k}f\left( j\right) j^{-s}(E_{j}^{s}-s!)
\end{equation*}%
\begin{equation*}
=\left\{ \Gamma \left( 2s+1\right) -\Gamma \left( s+1\right) ^{2}\right\} ^{%
\frac{1}{2}}\sum_{j=1}^{k}f\left( j\right) j^{-s}F_{j}^{(s)}\left( s\right) ,
\end{equation*}

\noindent and for a fixed $s\geq 1,$%
\begin{equation*}
F_{j}^{(s)}\left( s\right) =(E_{j}^{s}-s!)/\left\{ \Gamma \left( 2s+1\right)
-\Gamma \left( s+1\right) ^{2}\right\} ^{\frac{1}{2}}
\end{equation*}

\noindent is a sequence of independent mean zero random variables with
variance one. \noindent We have

\begin{eqnarray}
T_{n}(f,s)-\gamma ^{s}a_{n}(f,s)&=&\gamma ^{s}\left\{ \Gamma \left(
2s+1\right) -\Gamma \left( s+1\right) ^{2}\right\} \sum_{j=1}^{k}f\left(
j\right) j^{-s}F_{j}^{(s)}\left( s\right)\nonumber \\
 &+&\sum\limits_{j=1}^{k}f(j)\zeta
_{n,j}(s).  \label{etap01a}
\end{eqnarray}

\noindent By (\ref{eq01}),%
\begin{eqnarray*}
\left\vert \zeta _{n,j}(s)\right\vert &=&s\left\{ \frac{\gamma }{j}%
E_{j}+\theta _{n,j}\left( \left\vert O_{\mathbb{P}}(p_{n})\right\vert +\frac{%
E_{j}}{j}\left\vert O_{\mathbb{P}}(b_{n})\right\vert \right) \right\} ^{s-1}
\\
&\times &\left( \left\vert O_{\mathbb{P}}(p_{n})\right\vert +\frac{E_{j}}{j}%
\left\vert O_{\mathbb{P}}(b_{n})\right\vert \right)
\end{eqnarray*}

\noindent When $(K1)$ holds, Kolmogorov's Theorem on sums of centered random
variables ensures that 
\begin{equation*}
\left\{ \Gamma \left( 2s+1\right) -\Gamma \left( s+1\right) ^{2}\right\}
^{1/2}\sum_{j=1}^{k}f\left( j\right) j^{-s}F_{j}^{(s)}\left( s\right)
\end{equation*}%
converges to the $rv$%
\begin{equation*}
\left\{ \Gamma \left( 2s+1\right) -\Gamma \left( s+1\right) ^{2}\right\}
^{1/2}\sum_{j=1}^{+\infty }f\left( j\right) j^{-s}F_{j}^{(s)}\left( s\right)
=\left\{ \Gamma \left( 2s+1\right) -\Gamma \left( s+1\right) ^{2}\right\}
^{1/2}\mathcal{L}(f,s)
\end{equation*}

\noindent which is centered and has variance one, as completely described
in Lemma (\ref{lemma1}).\\

\bigskip \noindent  Now $(H0a)$ and Lemma (\ref{lemma2}) ensure that
the second term of (\ref{etap01a}) tends to zero in probability. Then if $%
a_{n}(f,s)\rightarrow \infty ,$ we get that $T_{n}(f,s)/a_{n}(f,s)%
\rightarrow \gamma ^{s}.$ This proves Part (A)(1) of Theorem (\ref{theo2})
for $\gamma >0$. Further if $(H2a)$ holds, Lemma (\ref{lemma2}) and the
first point yields that $T_{n}(f,s)-\gamma ^{s}a_{n}(f,s)$ asymptotically
behaves as $L(f,s)=\left\{ \Gamma \left( 2s+1\right) -\Gamma \left(
s+1\right) ^{2}\right\} ^{1/2}\sum_{j=1}^{+\infty }f\left( j\right)
j^{-s}F_{j}^{(s)}\left( s\right) $ since the second term of (\ref{etap01a})
is zero at infinity. This proves Part (A)(1) of Theorem (\ref{theo3}) for $%
\gamma >0$.\newline

\bigskip

\noindent Now suppose that $(K1)$ does not hold and that $(K2)$ and $(H2a)$
both hold. Also $(H2a)$ implies via Lemma (\ref{lemma1}) that the second
term, when divided by $\sigma _{n}(f,s),$ tends to zero in probability. Next
by Lemma (\ref{lemma2}), 
\begin{equation*}
\sigma _{n}^{-1}(f,s)\left\{ \Gamma \left( 2s+1\right) -\Gamma \left(
s+1\right) ^{2}\right\} ^{1/2}\sum_{j=1}^{k}f\left( j\right)
j^{-s}F_{j}^{(s)}\left( s\right)
\end{equation*}%
asymptotically behaves as a $\mathcal{N}(0,1)$ \textit{rv} under $(K2)$. \
It follows under these circumtances that%
\begin{equation*}
(a_{n}(f,s)/\sigma _{n}(f,s))(T_{n}(f,s)/a_{n}(f,s)-\gamma ^{s})\rightarrow
N(0,1).
\end{equation*}

\bigskip \noindent This ends the proof of Part (A)(2) of Theorem (\ref{theo2}). Further,
whenever $a_{n}(f,s)/\sigma _{n}(f,s)\rightarrow \infty ,$%
\begin{equation*}
T_{n}(f,s)/a_{n}(f,s)\rightarrow \gamma ^{s},
\end{equation*}

\noindent which establishes Part (A) (2) of Theorem (\ref{theo2}).\newline

\bigskip

\noindent Case B : $F\in D\left( G_{0}\right) ,\gamma =0$. Use
representations (\ref{rep3a}) and (\ref{rep3b}) to get for $1\leq j\leq
U_{k+1,n},$

\begin{equation*}
G^{-1}(1-U_{j,n})-G^{-1}(1-U_{j+1,n})=a\left( U_{j+1,n}\right) -a\left(
U_{j,n}\right) +\int\limits_{U_{j,n}}^{U_{j+1,n}}a(t)t^{-1}dt.
\end{equation*}%
Remark that for $U_{1,2}<u,v<U_{k,n}$

\begin{equation*}
\frac{v}{u}<\frac{U_{k,n}}{U_{1,n}}=\frac{U_{k,n}}{U_{k-1,n}}\times \frac{%
U_{k-1,n}}{U_{k-2,n}}\times \frac{U_{k-2,n}}{U_{k-3,n}}\times \cdot \cdot
\cdot \times \frac{U_{2,n}}{U_{1,n}}
\end{equation*}%
and

\begin{equation*}
0\leq \log \left( \frac{u}{v}\right) =\sum_{j=1}^{k-1}\log \left( \frac{%
U_{j+1,n}}{U_{j,n}}\right)
=\sum_{j=1}^{k-1}j^{-1}E_{j}=\sum_{j=1}^{k-1}j^{-1}\left( E_{j}-1\right)
+\sum_{j=1}^{k-1}j^{-1}.
\end{equation*}

\bigskip \noindent 
By Kolmogorov's theorem for partial sums of independent and mean zero random
variables, $\sum_{j=1}^{k-1}j^{-1}\left( E_{j}-1\right) $ converges in law
to a finite $rv$ $E$. We have $\sum_{j=1}^{k-1}j^{-1} \sim \log k$ and this
ensures that for $1\leq u\leq U_{k+1,n},$ $(1+p\left( u\right)
)/(1+p(U_{k+1,n}))-1=O_{P}(p_{n}),$ 
\begin{equation*}
\exp \left( \int_{u}^{1}b(t)t^{-1}dt\right) /\exp \left(
\int_{U_{k+1,n}}^{1}b(t)t^{-1}dt\right) -1=O_{P}(b_{n}\log k)
\end{equation*}%
both uniformly in $1\leq u\leq U_{k+1,n}$ and finally%
\begin{equation*}
a(U_{j+1,n})/a(U_{j,n})=(1+O_{P}(p_{n}))\exp (O_{P}(b_{n})E_{j}/j),
\end{equation*}

\bigskip \noindent 
But for $1\leq j\leq k,\left\vert j^{-1}E_{j}\right\vert \leq
\sum_{j=1}^{k-1}j^{-1}E_{j}=O_{p}(\log k).$ Then if $b_{n}\log k\rightarrow
_{\mathbb{P}}0,$

\begin{equation*}
a(U_{j+1,n})/a(U_{j,n}) =(1+O_{P}(p_{n}))(1+b_{n}E_{j}/j)
\end{equation*}
\begin{equation*}
=1+O_{P}(p_{n})+O_{P}(p_{n}b_{n}E_{j}/j)+O_{P}(b_{n}E_{j}/j).
\end{equation*}

\bigskip \
\noindent Finally, since $a(k/n)/a(U_{k+1,n})=1+O_{P}(1),$ it follows that

\begin{equation*}
a(u)/a(k/n)=1+O_{P}((p_{n}\vee b_{n})\log k)
\end{equation*}

uniformly in $1\leq u\leq U_{k+1,n}.$ This finally leads to for $1\leq j\leq
U_{k+1,n}$ ,

\begin{equation*}
B_{j,n}(s) =\left( G^{-1}(1-U_{j,n})-G^{-1}(1-U_{j+1,n})\right) /a(k/n)
\end{equation*}
\begin{equation*}
=(1+O_{P}((p_{n}\vee b_{n})\log k))\times \left\{
O_{P}(p_{n})+O_{P}(p_{n}b_{n}E_{j}/j)+O_{P}(b_{n}E_{j}/j)\right\}
\end{equation*}

\begin{equation*}
+\left\{ 1+O_{P}((p_{n}\vee b_{n})\log k)\right\} E_{j}/j
\end{equation*}

\begin{equation*}
=E_{j}/j+R_{j,n}(s),
\end{equation*}

\bigskip \noindent  where%
\begin{equation*}
R_{j,n}(s) =(1+O_{P}((p_{n}\vee b_{n})\log k))\times \left\{
O_{P}(p_{n})+O_{P}(p_{n}b_{n}E_{j}/j)+O_{P}(b_{n}E_{j}/j)\right\}
\end{equation*}

\begin{equation*}
+O_{P}((p_{n}\vee b_{n})\log k)E_{j}/j.
\end{equation*}

\noindent We can easily show that 
\begin{equation*}
R_{j,n}(s)=O_{P}(p_{n})+O_{P}((p_{n}\vee b_{n})\log k)E_{j}/j
\end{equation*}

\noindent and remark that $E_{j}/j=O_{P}(\log k)$ uniformly in $j\in
\{1,...,k\}$. This yields

\begin{equation*}
\frac{T_{n}\left( f,s\right) }{a^{s}\left( k/n\right) }=\sum_{j=1}^{k}f%
\left( j\right) \left\{ O_{P}(p_{n})+O_{P}((p_{n}\vee b_{n})\log
k)E_{j}/j\right\} ^{s}.
\end{equation*}

\noindent We have by the same methods used above

\begin{equation*}
\frac{T_{n}\left( f,s\right) }{a^{s}(k/n)}=\sum_{j=1}^{k}f\left( j\right)
j^{-s}E_{j}^{s}+\sum_{j=1}^{k}f\left( j\right) \xi _{n,j}(s)
\end{equation*}%
\begin{eqnarray*}
\xi _{n,j}(s) &=&s\left\{ E_{j}j^{-1}+\theta _{n,j}\left( O_{\mathbb{P}%
}(p_{n})+\frac{E_{j}}{j}O_{\mathbb{P}}(p_{n}\vee b_{n}\log k)\right)
\right\} ^{s-1} \\
&\times &\left( O_{\mathbb{P}}(p_{n})+\frac{E_{j}}{j}O_{\mathbb{P}%
}(p_{n}\vee b_{n}\log k)\right)
\end{eqnarray*}

\noindent And further 
\begin{equation*}
\frac{T_{n}\left( f,s\right) }{a^{s}\left( k/n\right) }-a_{n}\left(
f,s\right) =\sum_{j=1}^{k}f\left( j\right) j^{-s}\left( E_{j}^{s}-s!\right)
+\sum_{j=1}^{k}f\left( j\right) \xi _{n,j}(s).
\end{equation*}%
\begin{equation}
\frac{T_{n}\left( f,s\right) }{a^{s}\left( k/n\right) }=\left\{ \Gamma
\left( 2s+1\right) -\Gamma \left( s+1\right) ^{2}\right\}
^{1/2}\sum_{j=1}^{k}f\left( j\right) j^{-s}F_{j}^{(s)}\left( s\right)
+\sum_{j=1}^{k}f\left( j\right) \xi _{n,j}(s).  \label{etap01b}
\end{equation}

\noindent When we compare Formulas (\ref{etap01a}) and (\ref{etap01b}), $%
(H0a)-(H2a)-(H1a)$ with $(H0b)-(H2b)-(H1b)$, we use Lemmas (\ref{lemma1})
and (\ref{lemma2}) and reconduct almost the same conclusion already done for
the $\gamma >0$ case to prove the parts (B) of Theorems (\ref{theo2}) and (%
\ref{theo3}).

\section{Simulation Studies} \label{sec4}

Nowadays, simulation studies are very sophisticated and may be very
difficult to follow. Here, we want give a serious comparison of our
estimators with several analoges while keeping the study reasonably simple.
Let us begin to explain the stakes before proceeding any further. The
estimators of the extremal index generally use a number, say $k$ like in
this text, of the greatest observations : $X_{n-j+1}$, $1\leq j\leq k$. For
almost all such estimators, we have a small bias and a great variance for
large values of $k$, and the contrary happens for small values of $k$. This
leads to the sake of an optimal value of $k$ keeping both the bias and the
variance at a low level. A related method consists in considering a range of
values $kv(j)=kmin+j(kmin-kmin)/ksize$, $1 \leq j \leq ksize$ over which the observed values of the statistic are stable and well approximate the index. This second method seems preferable when comparing two estimators with respect to the bias.%
\newline

\bigskip \noindent  So we fix a sample size $n$ and consider the range of values as described above where $kmin$ and $kmax$ are suitably chosen. Thus checking the curves of two statistics over the interval $[mink,maxk]$ is a good tool for
comparing their performances. Next, for each $j$, $1\leq kmin\leq j\leq kmax$, we
compute the mean square error of the estimated values of $\gamma $ \ for
values of $k$ in a neighborhood of $kv(j)$, that is for $k \in [kv(j)-kstab,kv(j)+kstab]$, $1 \leq j \leq ksize$, where $kstab$ is also suitably fixed. The mimimun of these MES's certainly corresponds to the most stable zone and may be taken as the best estimation when it is low enough.\\

\bigskip \noindent Here, we compare our class of estimators, represented the optimal estimator (\ref{ODH}) and the boundary form
 (\ref{DH}) for $s\in[1,5]$, with the estimators of Hill and Dekkers \textit{et al.}. The estimators (\ref{ODH}) and (\ref{DH}) for $s\in[1,5]$ fall in the asymptotic normality area $s-1/2 \leq \tau$. In a larger study, we will include the Pickands' statistic
and consider the nongaussian asymptotic area.\newline

\bigskip \noindent The study will only cover the heavy tail case, that $\gamma>0$. The case $\gamma=0$ will be part of a large simulation paper. And we consider a pure Pareto law (I), and two perturbed ones laws (II) and (III):

\begin{table}[htbp]
\centering
\begin{tabular}{c}
(I) $F^{-1}(1-u)=u^{-\gamma}$ \\ 
(II) $F^{-1}(1-u)=u^{-\gamma}(1-u^{\beta })$ \\ 
(III) $F^{-1}(1-u)=u^{-\gamma}(1-(-1/\log u)^{\beta })$. \\ 
\end{tabular}%
\end{table}

\subsection{Simulations for $\protect\gamma>0$} Let $f(j)=j^\tau$, $\tau>0$. Our results say that $\left( \frac{T_{n}\left( \tau ,s\right) }{a_{n}\left( \tau ,s\right) }\right) ^{1/s}$ is an estimator of $\gamma \geq 0 $ if $s-1\leq \tau .$\newline

\bigskip \noindent Theoritically, we then have in hands an infinite class of estimators. We
should be able to find values of $s$ and $\tau $ leading a lowest
stable bias and hope that this bias will be lower than of the other
analogues, or to be at their order at least. We know from Subsection \ref{minimizationV}
that our class of estimators for $f=f_{\tau }$ has miminal asymptotic
variance for $s=\tau $. But it is not sure that this corresponds to the best
performance for finite samples. And following the remark of Deme \textit{et
al.} \cite{demedioplo} who noticed that the boundary case, that
discriminates the Gaussian and the non Gaussian asymptotic laws,  behaves
well, we include it also here, that is the case  $s-1/2=\tau $ . 

\noindent We fix the following values : $n=1000$, $kmin=105$, $kmax=375$, $ksize=100$, 
$kstab=5$. And we fix the number of replications to $B=1000$.\\

\noindent How to read our results? For each  $j \in [1, ksize]$, we compue the mean square error $MSE(j)$ when $k$ spans $[kv(j)-kstab,kv(j)+kstab]$, that is
$$
MSE(j)^2=\frac{1}{2kstab+1} \sum_{k \in [kv(j)-kstab,kv(j)+kstab]} (T_n(k)-\gamma)^2.
$$

\noindent Next we take the mimimum and the maximum values of these MSE(j)'s denoted as \textit{Min} and \textit{Max}. The difference (\textit{df}) Diff=\textit {Max}-\textit{Min} is reported as well as the middle term $Mid=(Min+Max)/2$.\\

\noindent We classify the estimators with respect to both the values of \textit{Mid} and \textit{Diff}. If the values \textit{Diff} are of the same order for two estimators, we will prefer the one with the minimum value of \textit{Mid}. That mean that this latter estimator is more stable and then, is better.\\

\noindent Since the Hill estimator and the Dekkers \textit{et al.} do not depend on parameters, we have conducted a series of simulations and the performances are of the order of values given in Table \ref{tabHD1}. Next, we conduct simulations on the performances of the boundary Double Hill
and the optimal double Hill statistics for $s=1$, ..., $s=5$ in Tables \ref{tabDH1}, \ref{tabODH1}, \ref{tabDH2}, \ref{tabODH2}, \ref{tabDH3}, \ref{tabODH3}.

\begin{center}
\begin{table}[tbp]
\begin{tabular}{|l||ll|ll|ll||}
\hline \hline
Values &    \multicolumn{2}{c}{ Model I}         &\multicolumn{2}{c}{ Model II}        &  \multicolumn{2}{c||}{ Model III}   \\ \hline
       &      Hill            & Dekkers          & Hill             & Dekkers          &  Hill             & Dekkers          \\ \hline
Min		 &     $1.5291 10^{-5}$ & $7.6835 10^{-3}$ & $3.5751 10^{-3}$ & $9.3245 10^{-5}$ & $1.0428 10^{-1}$  & $5.5751 10^{-2}$ \\  \hline
Max		 &     $2.0338 10^{-3}$ & $1.5158 10^{-2}$ & $3.7242 10^{-2}$ & $1.6484 10^{-2}$ & $8.4425 10^{-1}$  & $5.4568 10^{-1}$ \\   \hline
Diff	 &     $2.0185 10^{-3}$ & $7.4745 10^{-3}$ & $3.3667 10^{-2}$ & $1.6391 10^{-2}$ & $7.3996 10^{-1}$  & $4.9011 10^{-1}$ \\   \hline
Mid		 &     $1.0245 10^{-3}$ & $1.1420 10^{-2}$ & $2.0408 10^{-2}$ & $8.2887 10^{-3}$ & $4.7427 10^{-1}$  & $3.0080 10^{-1}$ \\ \hline \hline
\end{tabular}
\caption{Performances of Hill and Dekkers \textit{et al.} estimators}
\label{tabHD1}
\end{table}
\end{center}

\begin{center}
\begin{table}[tbp]
\begin{tabular}{|l||l|l|l|l|l||}
\hline \hline
Values &                               \multicolumn{5}{c||}{ Model I Double Hill}              \\ \hline
       &      s=1              & s=2               & s=3              & s=4               &  s=5            \\ \hline
Min		 & $6.2030 10^{-4}$      & $1.1113 10^{-3}$  & $2.5988 10^{-2}$ & $5.2379 10^{-3}$  & $6.6558 10^{-3}$  \\ \hline
Max		 & $1.9037 10^{-3}$      & $2.1814 10^{-3}$  & $3.0142 10^{-2}$ & $6.0324 10^{-3}$  & $8.4433 10^{-3}$  \\ \hline
Diff	 & $1.2834 10^{-3}$      & $1.0701 10^{-3}$  & $4.1537 10^{-3}$ & $7.9454 10^{-4}$  & $1.7874 10^{-3}$  \\ \hline
Mid		 & $1.2620 10^{-3}$      & $1.6463 10^{-3}$  & $2.8065 10^{-2}$ & $5.6352 10^{-3}$  & $7.5496 10^{-3}$  \\ \hline \hline
\end{tabular}
\caption{Performances of the Boundary Double Hill statistics for s=1,...,5 with Model I}
\label{tabDH1}
\end{table}
\end{center}

\begin{center}
\begin{table}[tbp]
\begin{tabular}{|l||l|l|l|l|l||}
\hline \hline
Values &   \multicolumn{5}{c||}{ Model I  Optimal Double Hill}                 \\\hline
       &      s=1         & s=2               & s=3                & s=4               &  s=5            \\ \hline
Min		 & $4.7128 10^{-3}$ & $1.4042 10^{-4}$  & $9.4369 10^{-7}$ & $5.4566 10^{-3}$  & $3.8495 10^{-3}$  \\ \hline
Max		 & $9.8514 10^{-3}$ & $5.3747 10^{-3}$  & $4.1170 10^{-3}$ & $1.0802 10^{-2}$  & $1.6068 10^{-2}$  \\ \hline
Diff	 & $5.1386 10^{-3}$ & $5.2342 10^{-3}$  & $4.1160 10^{-3}$ & $5.3454 10^{-3}$  & $1.2218 10^{-2}$  \\ \hline
Mid		 & $7.2821 10^{-3}$ & $2.7575 10^{-3}$  & $2.0589 10^{-3}$ & $8.1293 10^{-3}$  & $9.9589 10^{-3}$  \\ \hline \hline
\end{tabular}
\caption{Performances of the Optimal Double Hill statistics for s=1,...,5 with Model I}
\label{tabODH1}
\end{table}
\end{center}

\begin{center}
\begin{table}[tbp]
\begin{tabular}{|l||l|l|l|l|l||}
\hline \hline
Values &     \multicolumn{5}{c||}{ Model II Double Hill}          \\ \hline
       &      s=1              & s=2               & s=3              & s=4               &  s=5            \\ \hline
Min		 & $1.8258 10^{-3}$      & $1.3226 10^{-3}$  & $2.9811 10^{-2}$ & $1.5621 10^{-5}$  & $8.1246 10^{-6}$  \\ \hline
Max		 & $1.0097 10^{-2}$      & $9.6926 10^{-3}$  & $3.3995 10^{-2}$ & $5.2278 10^{-3}$  & $6.2451 10^{-3}$  \\ \hline
Diff	 & $8.2720 10^{-3}$      & $8.3699 10^{-3}$  & $4.1840 10^{-3}$ & $5.2122 10^{-3}$  & $6.2369 10^{-3}$  \\ \hline
Mid		 & $5.9618 10^{-3}$      & $5.5076 10^{-3}$  & $3.1903 10^{-2}$ & $2.6217 10^{-3}$  & $3.1266 10^{-3}$  \\ \hline \hline
\end{tabular}
\caption{Performances of the Boundary Double Hill statistics for s=1,...,5 with Model II}
\label{tabDH2}
\end{table}
\end{center}

\begin{center}
\begin{table}[tbp]
\begin{tabular}{|l||l|l|l|l|l||}
\hline \hline
Values &    \multicolumn{5}{c||}{ Model II  Optimal Double Hill}             \\ \hline
       &      s=1              & s=2               & s=3                & s=4               &  s=5            \\ \hline
Min		 & $1.8432 10^{-3}$      & $1.3451 10^{-3}$  & $1.4954 10^{-3}$   & $1.5621 10^{-5}$  & $8.1246 10^{-6}$  \\ \hline
Max		 & $1.0097 10^{-2}$      & $9.6926 10^{-3}$  & $3.3995 10^{-2}$   & $7.9972 10^{-3}$  & $6.1712 10^{-3}$  \\ \hline
Diff	 & $8.2546 10^{-3}$      & $8.3474 10^{-3}$  & $3.2499 10^{-2}$   & $7.9816 10^{-3}$  & $6.1630 10^{-3}$  \\ \hline
Mid		 & $5.9705 10^{-3}$      & $5.5188 10^{-3}$  & $1.7745 10^{-2}$   & $4.0064 10^{-3}$  & $3.0896 10^{-3}$  \\ \hline \hline
\end{tabular}
\caption{Performances of the Optimal Double Hill statistics for s=1,...,5 with Model II}
\label{tabODH2}
\end{table}
\end{center}

\begin{center}
\begin{table}[tbp]
\begin{tabular}{|l||l|l|l|l|l||}
\hline \hline
Values &   \multicolumn{5}{c||}{ Model III Double Hill}     \\  \hline
       &      s=1              & s=2               & s=3              & s=4               &  s=5            \\ \hline
Min		 & $3.3517 10^{-2}$      & $3.1446 10^{-2}$  & $3.1446 10^{-2}$ & $2.4937 10^{-2}$  & $2.5509 10^{-2}$  \\ \hline
Max		 & $1.8794 10^{-1}$      & $2.7110 10^{-1}$  & $2.7110 10^{-1}$ & $6.1901 10^{-1}$  & $9.5387 10^{-1}$   \\ \hline
Diff	 & $1.5442 10^{-1}$      & $2.3965 10^{-1}$  & $2.3965 10^{-1}$ & $5.9407 10^{-1}$  & $9.2836 10^{-1}$   \\ \hline
Mid		 & $1.1072 10^{-1}$      & $1.5112 10^{-1}$  & $1.5127 10^{-1}$ & $3.2197 10^{-1}$  & $4.8969 10^{-1}$   \\ \hline \hline
\end{tabular} 
\caption{Performances of the Boundary Double Hill statistics for s=1,...,5 with Model III}
\label{tabDH3}
\end{table}
\end{center}

\begin{center}
\begin{table}[tbp]
\begin{tabular}{|l||l|l|l|l|l||}
\hline \hline
Values &   \multicolumn{5}{c||}{ Model III Optimal Double Hill}       \\ \hline
       &      s=1              & s=2               & s=3                & s=4               &  s=5            \\ \hline 
Min		 & $3.1979 10^{-2}$      & $3.1821 10^{-2}$  & $3.1821 10^{-2}$   & $2.5361 10^{-2}$  & $2.6060 10^{-2}$ \\ \hline
Max		 & $1.8792 10^{-1}$      & $2.7110 10^{-1}$  & $2.7110 10^{-1}$   & $6.1901 10^{-1}$  & $9.5387 10^{-1}$ \\ \hline
Diff	 & $1.5594 10^{-1}$      & $2.3928 10^{-1}$  & $2.3928 10^{-1}$   & $5.9364 10^{-1}$  & $9.2781 10^{-1}$ \\ \hline 
Mid		 & $1.0995 10^{-1}$      & $1.5146 10^{-1}$  & $1.5146 10^{-1}$   & $3.2218 10^{-1}$  & $4.8996 10^{-1}$   \\ \hline \hline
\end{tabular}
\caption{Performances of the Optimal Double Hill statistics for s=1,...,5 with Model III}
\label{tabODH3}
\end{table}
\end{center}

\noindent Now, we are able to draw a number of conclusions and remarks based on the Tables \ref{tabHD1}, \ref{tabDH1}, \ref{tabODH1}, \ref{tabDH2}, \ref{tabODH2}, \ref{tabDH3}, \ref{tabODH3}.\\

\bigskip \noindent \textbf{(1.)} Model I : Compared to Hill's statistic and Dekkers \textit{et al.}'s estimator, our Double Hill and Optimal Double Hill estimors are generally more stable and mostly better with regard to the middle value for $s=1$, $s=2$, $s=4$.\\

\bigskip \noindent \textbf{(2.)} Model II : we have simular results.\\

\bigskip \noindent \textbf{(3.)} Model III : For $s\leq3$. We also get similar results.\\

\bigskip \noindent \textbf{(4.)} As a general conclusion, we say that our Double Hill estimator, for both cases of of boundary and minimum variance, behaves like these Hill's and Dekkers \textit{et al.}'s estimator, are more stable and sloghtly better.\\

\bigskip \noindent \textbf{(5.)} All these estimators present poorer performances for the very perturbated model III. But unlikely to the Hill's and Dekkers \textit{et al.}'s estimators, we have in hand a SPMEEXI and hopefully we will be able to get better new estimators for model III through suitable combinations in future works. This is important since the Hill's and Dekkers \textit{et al.}'s are the most used in the applications.\\

\bigskip \noindent \textbf{(6.)} We may find theses patterns on Figure 1 and Figure 2.\\

\begin{figure}[htbp]
	\centering
		\includegraphics[width=0.89\textwidth]{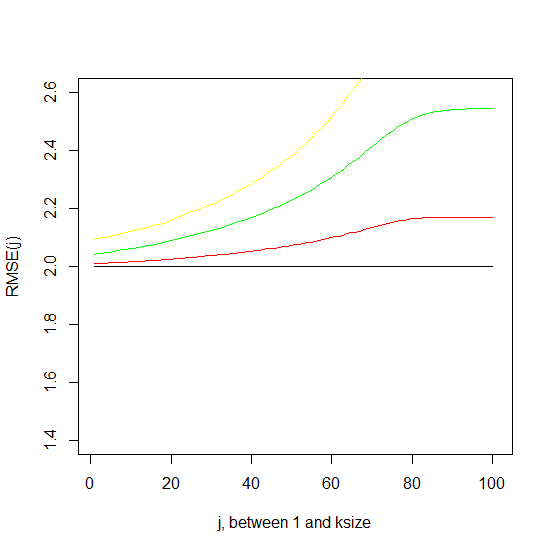}
	\label{fig1}
	\caption{Figure 1 : curves of the mean values of the RMSE's at 100 values of k(j) (j=1,...,100) computed on eleven points around k(j) for the statistics : Hill [blue], Dekkers \textit{et al.} [green], Boundary Double Hill [red] and Optimal Double Hill [yellow] for s=1}
\end{figure}

\begin{figure}[htbp]
	\centering
		\includegraphics[width=0.89\textwidth]{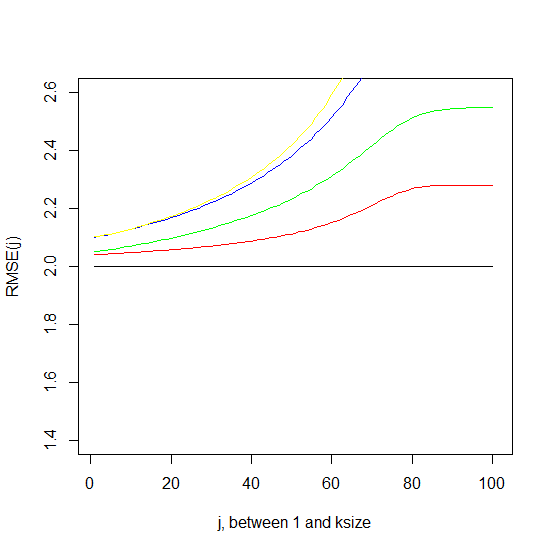}
	\label{fig2}
	\caption{Figure 2 : curves of the mean values of the RMSE's at 100 values of k(j) (j=1,...,100) computed on eleven points around k(j) for the statistics : Hill [blue], Dekkers \textit{et al.} [green], Boundary Double Hill [red] and Optimal Double Hill [yellow] for s=2}
\end{figure}

\section{Technical results} \label{sec5}

\label{sec6} \label{sectiontool}

\subsection{Technical lemmas}

We begin by this simple lemma where we suppose that we are given a sequence
of independent and identically mean zero $rv^{\prime }s$ $%
F_{1},F_{2},...,F_{k}$ with variance unity. Denote 
\begin{equation*}
A(f,s)=\sum_{j=1}^{\infty }f(j)^{2}j^{-2s}
\end{equation*}%
and 
\begin{equation*}
\sigma _{n}^{2}(f,s)=\sum_{j=1}^{k}f(j)^{2}j^{-2s}
\end{equation*}

\begin{lemma}
\label{lemmatool} \label{lemma1} Let 
\begin{equation*}
V_{n}(f,s)=\sum_{j=1}^{k}f(j)j^{-s}(E_{j}^{s}-s!)=:\left\{ \Gamma \left(
2s+1\right) -\Gamma \left( s+1\right) ^{2}\right\}
^{1/2}\sum_{j=1}^{k}f(j)j^{-s}F_{j}^{(s)},
\end{equation*}%
wher the $F_{j}^{(s)}$ are independent centered random variables with
variance one.

(1) If $(K1)$ \ : $A(f,s)<\infty $, then 
\begin{equation*}
V_{n}(f,s)\leadsto \left\{ \Gamma \left( 2s+1\right) -\Gamma \left(
s+1\right) ^{2}\right\} ^{1/2}\mathcal{L}(f,s).
\end{equation*}
\end{lemma}

(2) If $(K2)$ : $B_{n}(f,s)=\max \{f(j)j^{-s}/\sigma _{n}(f,s),1\leq j\leq
k\}\rightarrow 0,$ then 
\begin{equation*}
\sigma _{n}^{-1}(f,s)V_{n}(f,s)\leadsto \mathcal{N}(0,1)
\end{equation*}

\begin{proof}
Put 
\begin{equation*}
V_{n}^{\ast }(f,s)=\sigma _{n}(f,s)^{-1}V_{n}(f,s).
\end{equation*}%
and suppose that $(K1)$ holds. Then Kolmogorov's Theorem for sums of zero
mean indenpendent rv's applies. Since the series $\sum_{j\geq
1}Var(f(j)j^{-s}F_{j}^{(s)})$ is finite, we have : 
\begin{equation*}
V_{n}(f,s)\rightarrow \left\{ \Gamma \left( 2s+1\right) -\Gamma \left(
s+1\right) ^{2}\right\} ^{-1/2}\sum_{j=1}^{\infty
}f(j)j^{-s}F_{j}^{(s)}=\left\{ \Gamma \left( 2s+1\right) -\Gamma \left(
s+1\right) ^{2}\right\} ^{1/2}\mathcal{L}(f,s),
\end{equation*}

\noindent Now suppose that $(K2)$ holds. Let us evaluate the moment
generating function of $V_{n}^{\ast \ast }(f,s)=\sigma
_{n}(f,s)^{-1}\sum_{j=1}^{k}f(j)j^{-s}F_{j}^{(s)}:$

\begin{equation}
\phi _{V_{n}^{\ast \ast }(f,s)}(t)=\prod_{j=1}^{k}\phi
_{F_{j}^{(s)}}(tf(j)j^{-s}\sigma _{n}(f,s)^{-1}).  \label{tfl2}
\end{equation}%
We use the expansion common characteristic function $\phi (u)=\phi
_{F_{j}^{(s)}}(u)=1-u^{2}/2+u^{2}\varepsilon (v),$ where $\varepsilon
(v)\rightarrow 0,$ uniformly in $\left\vert u\right\vert \leq v\rightarrow
0. $ Remind that for $1\leq j\leq k,$ $0\leq f(j)j^{-s}\sigma
_{n}(f,s)^{-1}\leq B_{n}(f,s)\rightarrow 0$ and then 
\begin{equation*}
\phi _{V_{n}^{\ast \ast
}(f,s)}(t)=\prod_{j=1}^{k}(1-t^{2}f(j)^{2}j^{-2s}\sigma
_{n}^{-2}(f,s)/2+t^{2}f(j)^{2}j^{-2s}\sigma _{n}^{2}(f,s)\varepsilon (B_{n}))
\end{equation*}%
\begin{equation*}
=\exp (\sum_{j=1}^{k}\log (1-t^{2}f(j)^{2}j^{-2s}\sigma
_{n}^{2}(f,s)/2+t^{2}f(j)^{2}j^{-2s}\sigma _{n}^{2}(f,s)\varepsilon
(B_{n}))).
\end{equation*}%
By using a first order expansion the logarithmic function in the
neighborhood of unity, we have 
\begin{equation*}
\phi _{V_{n}^{\ast }(f)}(t)=\exp (\sum_{j=1}^{k}-t^{2}f(j)^{2}j^{-2s}\sigma
_{n}(f)^{-2}/2+t^{2}f(j)^{2}j^{-2s}\sigma _{n}^{-2}(f)\varepsilon (B_{n})),
\end{equation*}%
where the function $\varepsilon (B_{n})$ may change from one line to an
other, but always tends to zero. Hence 
\begin{equation*}
\phi _{V_{n}^{\ast \ast }(f,s)}(t)=\exp
(\sum_{j=1}^{k}-t^{2}/2+t^{2}\varepsilon (B_{n}))\rightarrow \exp (-t^{2}/2).
\end{equation*}%
and 
\begin{equation*}
V_{n}^{\ast \ast }(f)\rightarrow \mathcal{N}(0,1).
\end{equation*}%
and then%
\begin{equation*}
V_{n}^{\ast }(f)\rightarrow \mathcal{N}(0,\left\{ \Gamma \left( 2s+1\right)
-\Gamma \left( s+1\right) ^{2}\right\} ).
\end{equation*}
\end{proof}

\begin{lemma}
\label{lemma2} If any of $(H1a)$, $(H1b)$, $(H2a)$ and $(H2b)$ holds with an
abitrary $\lambda >1$, then their analogues where $b_{n}(\lambda )$ is
replaced with $b_{n}$\ and $p_{n}(\lambda )$ is replaced with $p_{n}$ also
hold.
\end{lemma}

\noindent Proof. We have to prove this only for one case. The others are
similarly done. We begin to recall the Balkema result, that is $\sqrt{n}%
((n/k)U_{k,n}-1)\rightarrow _{d}N(0,1)$ which entails that $\sqrt{n}%
((n/k)U_{k+1,n}-1)\rightarrow _{d}N(0,1)$ and next $(n/k)U_{k+1,n}%
\rightarrow _{\mathbb{P}}1.$ Then for any $\varepsilon >0,$ for any $\lambda
>1,$ we have for large values of $n$, say $n\geq n_{1},$%
\begin{equation*}
\mathbb{P}(U_{k+1,n}>\lambda n/k)\leq \varepsilon /3.
\end{equation*}%

\noindent Next by the definition of \ $V_{n}=O_{p}(p_{n})$ and $W_{n}=O_{p}(b_{n}),$ there
exists $C_{0}$ such for large values of $n,$ say $n\geq n_{2},$%
\begin{equation*}
\mathbb{P}(\left\vert V_{n}\right\vert >C_{0}p_{n})\leq \varepsilon
/3,P(\left\vert W_{n}\right\vert >C_{0}b_{n})\leq \varepsilon /3.
\end{equation*}%
Then for $n\geq \max (n_{1},n_{2}).$\ Recall that $b_{n}(\lambda
)=\{\left\vert b(t)\right\vert ,t\leq \lambda k/n\}$, $p_{n}(\lambda
)=\{\left\vert p(t)\right\vert ,t\leq \lambda k/n\}$, $b_{n}=\{\left\vert
b(t)\right\vert ,t\leq U_{k+1,n}\}$ and $p_{n}=\{\left\vert p(t)\right\vert
,t\leq U_{k+1,n}\}.$ We have%
\begin{equation*}
\mathbb{P}(\left\vert V_{n}\right\vert \leq C_{0}p_{n},\left\vert
W_{n}\right\vert \leq C_{0}b_{n},b_{n}\leq b_{n}(\lambda ),p_{n}\leq
p_{n}(\lambda ))\geq 1-\varepsilon .
\end{equation*}%
And next, since $s\geq 1,$ 
\begin{equation*}
\mathbb{P}\left\vert s\left\{ \frac{\gamma }{j}E_{j}+\theta _{n,j}\left( O_{%
\mathbb{P}}(p_{n})+\frac{E_{j}}{j}O_{\mathbb{P}}(b_{n})\right) \right\}
^{s-1}\left( O_{\mathbb{P}}(p_{n})+\frac{E_{j}}{j}O_{\mathbb{P}%
}(b_{n})\right) \right\vert
\end{equation*}%
\begin{equation*}
\left. \leq s\left\vert \frac{\gamma }{j}E_{j}+\left( C_{0}p_{n}(\lambda )+%
\frac{E_{j}}{j}C_{0}b_{n}(\lambda )\right) \right\vert ^{s-1}\left(
C_{0}p_{n}(\lambda )+\frac{E_{j}}{j}C_{0}b_{n}(\lambda )\right) \right) \geq
1-\varepsilon
\end{equation*}%
Suppose that for an arbitrary $\lambda >1,$ 
\begin{equation*}
c_{n}(\lambda )=s\left\vert \frac{\gamma }{j}E_{j}+\left( C_{0}b_{n}(\lambda
)+\frac{E_{j}}{j}C_{0}b_{n}(\lambda )\right) \right\vert ^{s-1}\left(
C_{0}p_{n}(\lambda )+\frac{E_{j}}{j}C_{0}b_{n}(\lambda )\right) \rightarrow
_{\mathbb{P}}0
\end{equation*}%
and put 
\begin{equation*}
c_{n}=\left\vert s\left\{ \frac{\gamma }{j}E_{j}+\theta _{n,j}\left( O_{%
\mathbb{P}}(p_{n})+\frac{E_{j}}{j}O_{\mathbb{P}}(b_{n})\right) \right\}
^{s-1}\left( O_{\mathbb{P}}(p_{n})+\frac{E_{j}}{j}O_{\mathbb{P}%
}(b_{n})\right) \right\vert
\end{equation*}%
$,$ we have for any $\eta >0,$ and a fixed $\lambda >1$ and for large values
of $n.$ This gives 
\begin{equation*}
\mathbb{P}(c_{n}>\eta )=P((c_{n}>\eta )\cap (c_{n}<c_{n}(\lambda )))+\mathbb{%
P}((c_{n}>\eta )\cap (c_{n}\geq c_{n}(\lambda )))
\end{equation*}%
\begin{equation*}
\leq (c_{n}(\lambda )>\eta )+\varepsilon .
\end{equation*}%
Now letting $n\rightarrow +\infty ,$ we have 
\begin{equation*}
\lim \sup_{n\rightarrow \infty }\mathbb{P}(c_{n}>\eta )\leq \varepsilon .
\end{equation*}%
By letting $\varepsilon \downarrow 0,$ one achieves the proof, that is $%
c_{n}\rightarrow _{\mathbb{P}}0.$

\subsection{Integral computations}

Let $b\geq 1,$ we get by comparing the area under the curve $x\longmapsto
x^{-b}$ from $j$ to $k-1$ and those of the rectangles based on the intervals 
$[h,h+1],$ $h=j,..,k-2,$ we get%
\begin{equation*}
\sum_{h=j+1}^{k-1}h^{-b}\leq \int_{j}^{k-1}x^{-b}dx\leq
\sum_{h=j}^{k-2}h^{-b},
\end{equation*}%
that is%
\begin{equation}
\int_{j}^{k-1}x^{-b}dx+(k-1)^{-b}\leq \sum_{h=j}^{k-1}h^{-b}\leq
\int_{j}^{k-1}x^{-b}dx+j^{-b}.  \label{integ0}
\end{equation}%
For $b=1$, we get 
\begin{equation}
\frac{1}{k-1}\leq (\sum_{h=j}^{k-1}\frac{1}{h})-\log ((k-1)/j)\leq \frac{1}{j%
}.  \label{integ1}
\end{equation}%
For $b=2,$, we have 
\begin{equation*}
\frac{1}{j}-\frac{1}{k-1}+\frac{1}{(k-1)^{2}}\leq \sum_{h=j}^{k-1}h^{-2}\leq 
\frac{1}{j}-\frac{1}{k-1}+\frac{1}{j^{2}},  \label{integ2}
\end{equation*}%
that is%
\begin{equation*}
\frac{1}{(k-1)^{2}}\leq \sum_{h=j}^{k-1}h^{-2}-\frac{1}{j}(1-\frac{j}{k-1}%
)\leq \frac{1}{j^{2}}.
\end{equation*}%
As well, we have for $b>0,$%
\begin{equation*}
\sum_{h=j}^{k-2}h^{b}\leq \int_{j}^{k-1}x^{b}dx\leq \sum_{h=j+1}^{k-1}h^{b}
\end{equation*}%
and then%
\begin{equation}
\frac{1}{b+1}((k-1)^{b+1}-j^{b+1})+j^{b}\leq \sum_{h=j}^{k-1}h^{b}\leq \frac{%
1}{b+1}((k-1)^{b+1}-j^{b+1})+(k-1)^{b}.  \label{integ3}
\end{equation}%
Hence for $j$ fixed and $k\rightarrow \infty ,$ we get $%
\sum_{h=j}^{k-1}h^{b} $ $\sim (k-1)^{b+1}/(b+1).$\bigskip

\subsection{Computation of J(a,b,c,s)}

Recall%
\begin{equation*}
J(a,b,c,s)=\int_{0}^{\infty }(a+cx)^{s}(a+bx)e^{-x}dx\text{ and }%
I(a,b,s)=\int_{0}^{\infty }(a+bx)^{s}e^{-x}dx.
\end{equation*}

\subsubsection{Computation of I(a,b,s).}

We have by integration by parts%
\begin{equation*}
I(a,b,s)=\int_{0}^{\infty }(a+bx)^{s}e^{-x}dx=\left[ -e^{-x}(a+bx)^{s}\right]
_{0}^{\infty }+bs\int_{0}^{\infty }(a+bx)^{s-1}e^{-x}dx,
\end{equation*}%
that is, for any $s\geq 1,$%
\begin{equation*}
I(a,b,s)=a^{s}+sbI(a,b,s-1).
\end{equation*}%
By induction for $s\geq 1,$this leads to%
\begin{equation*}
I(a,b,s)=s!\sum_{h=0}^{s}b^{h}a^{s-h}/(s-h)!.
\end{equation*}

\subsubsection{Computation of J(a,b,c,s)}

We have integration by parts%
\begin{equation*}
J(a,b,c,s)=a^{s+1}+bI(a,c,s)+csJ(a,b,c,s-1).
\end{equation*}%
We get by induction for $\ell \geq 1,$%
\begin{eqnarray*}
J(a,b,c,s) &=&s!a\sum_{h=0}^{\ell }c^{h}a^{s-h}/(s-h)!+s!b\sum_{h=0}^{\ell
}c^{h}I(a,c,s-h)/(s-h)! \\
&+&s!c^{\ell +1}J(a,b,c,s-\ell -1)/(s-\ell -1)!
\end{eqnarray*}%
For $\ell +1=s,$ we arrive at%
\begin{eqnarray*}
J(a,b,c,s)
&=&s!a\sum_{h=0}^{s-1}c^{h}a^{s-h}/(s-h)!+s!b%
\sum_{h=0}^{s-1}c^{h}I(a,c,s-h)/(s-h)! \\
&+&s!c^{s}J(a,b,c,0).
\end{eqnarray*}%
Since $J(a,b,c,0)=I(a,b,1),$ we finally get%
\begin{eqnarray*}
J(a,b,c,s)
&=&s!a\sum_{h=0}^{s-1}c^{h}a^{s-h}/(s-h)!+s!b%
\sum_{h=0}^{s-1}c^{h}I(a,c,s-h)/(s-h)! \\
&+&s!c^{s}I(a,b,1)
\end{eqnarray*}

\subsection{Minimization of the asymptotic variance}

\label{minimizationV}

\bigskip For $s \geq 1$ fixed, We have to maximize 
\begin{equation*}
\ V_{n}(\tau ,s)=\frac{d_{n}(\tau ,s)}{\sigma _{n}(\tau ,s)}\times \frac{%
s\gamma ^{-1}}{s!\sqrt{\Gamma (2s+1)-\Gamma (s+1)^{2}}}.
\end{equation*}

\noindent with respect of $\tau > s-1/2$ for $s$, where $d_{n}(\tau
,s)=\sum_{j=1}^{k}j^{\tau -s}$ . We denote \ $q(s)=\frac{s\gamma ^{-1}}{s!%
\sqrt{\Gamma (2s+1)-\Gamma (s+1)^{2}}}.$ Let us find critical points. It is
easy to see that 
\begin{equation*}
\frac{\partial V_{n}(\tau ,s)}{\partial \tau }=\frac{d_{n}^{\prime }(\tau
,s)\sigma _{n}(\tau ,s)-d_{n}(\tau ,s)\sigma _{n}^{\prime }(\tau ,s)}{\left(
\sigma _{n}(\tau ,s)\right) ^{2}}\times q\left( s\right) .
\end{equation*}

\noindent A zero-point of $\frac{\partial V_{n}(\tau ,s)}{\partial \tau }$
obviously, is a solution of the ordinary differential equation. 
\begin{equation*}
\frac{d_{n}^{\prime }(\tau ,s)}{d_{n}(\tau ,s)}=\frac{\sigma _{n}^{\prime
}(\tau ,s)}{\sigma _{n}(\tau ,s)}.  \label{edo}
\end{equation*}

\noindent Its general solution is given by

\begin{equation}
\log d_{n}(\tau ,s)=\log \sigma _{n}(\tau ,s)+C(s).  \label{solg}
\end{equation}

\noindent By taking the particular value of $\tau =s$, we find that \ $%
C(s)=(1/2)\log k,$and (\ref{solg}) becomes 
\begin{equation*}
\sum_{j=1}^{k}j^{\tau -s}=\left( k\sum_{j=1}^{k}j^{2\left( \tau -s\right)
}\right) ^{1/2}.  \label{solg1}
\end{equation*}

\noindent This is the equality form of Cauchy-Schwarz's inequality with
respect to the usual scalar product in $\mathbb{R}^{k}$. Then  there exists
\ a constant $\lambda (s)$ such that \ $j^{\tau -s}=\lambda (s)$ for $1\leq
j\leq k.$ The only solution \ is $\tau =s.$ Now to show that $\tau $ is the
global maximum point, it suffices to notice that%
\begin{equation*}
\frac{1}{s!q\left( s\right) }\frac{\partial ^{2}V_{n}(\tau ,s)}{\partial
^{2}\tau }=\frac{1}{k\sqrt{k}}\left( \left( \sum_{j=1}^{k}\log j\right)
^{2}-k\sum_{j=1}^{k}\left( \log j\right) ^{2}\right) <0.
\end{equation*}

\begin{equation*}
\frac{1}{s!q\left( s\right) \sqrt(k)}\frac{\partial ^{2}V_{n}(\tau ,s)}{%
\partial ^{2}\tau }= -\left( \left( k^{-1}\sum_{j=1}^{k}\left( \log j\right)
^{2} _ (k^{-1})\sum_{j=1}^{k}\log j\right) ^{2} \right) <0.
\end{equation*}

\noindent since the left member is the opposite of a the empirical variance
of $\log j$, $1\leq j \leq k$. We conclude that the point $\tau=s$ is the
unique local miximum point. Then the global maximum is reached at $\tau=s$.

\end{document}